\documentclass[11pt]{article} 

\usepackage[utf8]{inputenc} 
\pdfoutput=1

\usepackage{geometry} 
\usepackage{amssymb}
\usepackage{amsmath}
\usepackage{appendix}
\usepackage{graphicx} 

\usepackage{booktabs} 
\usepackage{array} 
\usepackage{paralist} 
\usepackage{verbatim} 
\usepackage{subfig} 
\usepackage{gensymb}

\usepackage{fancyhdr} 
\usepackage{sectsty}
\allsectionsfont{\sffamily\mdseries\upshape} 

\usepackage[nottoc,notlof,notlot]{tocbibind} 
\usepackage[titles,subfigure]{tocloft} 

\title{On the Possibility of Large Axion Moduli Spaces}
\author{Tom Rudelius\\ \\
  \small Jefferson Physical Laboratory, Harvard University,\\
\small  Cambridge, MA 02138, USA\\
  }

\begin{document}
\setlength{\baselineskip}{16pt}
\begin{titlepage}
\maketitle
\begin{picture}(0,0)(0,0)
\put(325,300){hep-th/0202056}
\put(325,315){PUPT--2022}
\end{picture}
\vspace{-36pt}
\begin{abstract}
We study the diameters of axion moduli spaces, focusing primarily on type IIB compactifications on Calabi-Yau three-folds.  In this case, we derive a stringent bound on the diameter in the large volume region of parameter space for Calabi-Yaus with simplicial K\"{a}hler cone.  This bound can be violated by Calabi-Yaus with non-simplicial K\"{a}hler cones, but additional contributions are introduced to the effective action which can restrict the field range accessible to the axions.  We perform a statistical analysis of simulated moduli spaces, finding in all cases that these additional contributions restrict the diameter so that these moduli spaces are no more likely to yield successful inflation than those with simplicial K\"{a}hler cone or with far fewer axions.  Further heuristic arguments for axions in other corners of the duality web suggest that the difficulty observed in \cite{Banks} of finding an axion decay constant parametrically larger than $M_p$ applies not only to individual axions, but to the diagonals of axion moduli space as well.  This observation is shown to follow from the weak gravity conjecture of \cite{wgc}, so it likely applies not only to axions in string theory, but also to axions in any consistent theory of quantum gravity.

\end{abstract}
\thispagestyle{empty}
\setcounter{page}{0}
\end{titlepage}

\section{Introduction}

The inflationary paradigm \cite{guth, linde, albrecht&steinhardt} has become the predominant solution to the problems facing the standard model of cosmology.  Although recent measurements of B-mode polarization by the BICEP2 collaboration \cite{bicep2} can be explained largely by dust \cite{Planckdust, PlanckBicep}, a primordial tensor-to-scalar ratio $r \gtrsim 0.05$ is not ruled out and would likely be detected by the forthcoming generation of CMB polarization experiments.  If inflation driven by a scalar field is responsible for producing these tensor modes, the Lyth bound \cite{lyth} indicates that the scalar field must traverse a super-Planckian distance in field space during the course of its slow-roll evolution,
\begin{equation}
\Delta\phi \gtrsim \left( \frac{r}{0.01} \right)^{1/2}\, M_p
\label{lythbound}
\end{equation}
Here and henceforth, $M_p$ is the reduced Planck mass $\approx 2 \times 10^{18}$ GeV.  Maintaining the flat potential needed for slow-roll inflation is exceedingly difficult over a field range larger than $M_p$, as quantum corrections to the potential are naturally significant on Planckian scales.  Axions could potentially solve this problem, as their shift symmetry protects the inflationary potential from perturbative corrections, while instanton corrections break the shift symmetry to a discrete subgroup and generate a smooth, periodic potential like the one used in models natural inflation \cite{Freese}.  Furthermore, axions are ubiquitous in string compactifications and so give hope for an ultraviolet completion.

Achieving the prequisite number of $\approx 60$ $e$-foldings of inflation requires an axion decay constant just slightly larger than $M_p$, though an axion decay constant smaller than about $5 M_p$ is in tension with measurements of the spectral index $n_s$ by the \emph{Planck} collaboration \cite{Planck}.  However, in \cite{Banks}, Banks, Dine, Fox, and Gorbatov studied axions arising in various corners of the string duality web and found no instances of super-Planckian decay constants.  In all situations considered, decay constants much larger than $M_p$ were either impossible to achieve or else higher harmonics of the potential became relevant on scales $\phi \sim \mathcal{O}(M_p)$.  As a result, none of these axions make good inflaton candidates.

In \cite{Nflation}, a simple solution to this problem was proposed based on previous work in \cite{assisted1, assisted2, assisted3, assisted4, assisted5, assisted6, assisted7, assisted8, assisted9,assisted10}: use multiple axions, and travel along the diagonal direction, leading to a parametrically large field range relative to inflating along only one axion direction.  Nevertheless, several papers have raised the question of whether or not such a scenario is feasible.  In \cite{0710.3429}, the issue of moduli stabilization was considered, and it was found to be exceedingly difficult to stablize the K\"{a}hler moduli in a string compactification while leaving the axions light enough to inflate.  However, the authors did succeed in producing successful moduli stabilization in supergravity and concluded that the search for such a scenario in string theory is an open problem.   In \cite{CicoliDutta}, an explicit embedding of $N$-flation into string theory was proposed, in which K\"{a}hler moduli are stabilized in the LARGE Volume Scenario (LVS) of \cite{LVSpaper} by perturbative effects while non-perturbative effects are subdominant, leaving a certain subset of axions hierarchically lighter than the remaining scalar fields.  It is worth noting that for our purposes, the `large volume region' of parameter space is simply the regime in which the volumes of the curves of the compactification Calabi-Yau are $\gtrsim 1$ in string units, whereas the LVS assumes that such volumes are $\gg 1$.  For a related work on inflation in the LVS, see \cite{Quevedo}.

We will ignore such issues of moduli stabilization in this present work, concentrating instead on the outstanding geometric issues regarding the diameters of axion moduli spaces.  More specifically, we assume that the masses of the moduli are fixed to be large enough so that these fields decouple from inflationary dynamics.  In \cite{Nflation}, it was noted that the overall volume of a Calabi-Yau (in string units) must not grow linearly with $N$, or else the Planck mass would grow as $\sqrt{N}$, cancelling the Pythagorean gain from traveling along the diagonal.  However, it was asserted that suitable cancellations may occur, maintaining a small volume and allowing for parametric enhancement of moduli space diameter.   In \cite{Grimm}, this cancellation was called into question, and it was argued that in the large volume limit, axion decay constants should scale as $1/\sqrt{N}$, cancelling the Pythagorean gain from traveling along the diagonal.  Nevertheless, it was further suggested that a Calabi-Yau compactification with 2-cycles smaller than the string length squared may admit a parametrically large diameter of axion moduli space.  In particular, it was argued that the worldsheet instanton effects that arise from strings wrapping holomorphic curves generate positive corrections to the metric in (\ref{bmetric}), so that smaller cycles can result in larger distances.  This scenario, however, crucially depended on details of the toy geometry under investigation (in particular, an abundance of negative intersection numbers), which we shall soon see is unrealistic.

In \cite{kineticalignment}, the phenomenon of `kinetic alignment' was noted and suggested as a way of acheiving a parametrically-large diameter of axion moduli space, even with a broad distribution of axion decay constants.  In particular, given a metric on axion moduli space consisting of eigenvectors of random eigenvalue and direction, it was pointed out that for large $N$, the longest eigenvector almost always points along one of the diagonals of the space.  If one could achieve $\mathcal{O}(N^0)$ scaling of this largest eigenvalue, one would thereby attain parametric scaling of the moduli space diameter as $N^{1/2}$.  However, no explicit construction realizing this proposal has been produced.

Therefore, despite recent arguments for and against, it is still unknown whether or not one can achieve a parametrically-large axion moduli space in a region of IIB parameter space under theoretical control.  In this paper, we will address this issue and find evidence to the contrary. For a compactification geometry with a simplicial K\"{a}hler cone (a term to be defined in the next section and discussed more fully in the appendix), we derive a bound showing that parametric enhancement of the radius (half the diameter) of axion moduli space is impossible, and in fact is bounded by,
\begin{equation}
r \lesssim \left( \frac{15}{4} \right)^{1/2} \pi \, M_p.
\label{firstbound}
\end{equation}
Note that this bound is independent of $N$ and is, in fact, independent of the compactification geometry altogether (aside from the simpliciality of the K\"{a}hler cone).  Further, in all studied examples, we find an even more stringent bound,
\begin{equation}
r \lesssim \left( \frac{3}{4} \right)^{1/2} \pi\, M_p.
\label{secondbound}
\end{equation}
For a Calabi-Yau with a non-simplicial K\"{a}hler cone, the derivation of this bound breaks down in a rather interesting way, and we are left to search for experimental evidence in the landscape of possible Calabi-Yau compactifications.  Among a broad class of simulated Calabi-Yau moduli spaces, we find compelling evidence that both (\ref{firstbound}) and (\ref{secondbound}) are indeed violated, so that there likely exist Calabi-Yau manifolds with regions of moduli space under theoretical control admitting a radius of order $10\, M_p$.  However, we find no evidence that metric eigenvalues should scale better than $1/N$, so the Pythagorean gain $\sqrt{N}$ from traveling along the diagonal is offset by the $1/\sqrt{N}$ scaling of distances.

Furthermore, although the non-simpliciality of the K\"{a}hler cone suffices to avoid the bound (\ref{rbound}), it introduces additional contributions to the potential from objects wrapping the additional curves.  We find in \emph{all} cases considered that these additional contributions play a significant role whenever a metric eigenvalue becomes large, restricting the range over which an axion can travel during inflation to less than $\pi M_p$.  This is quite reminiscent of the additional harmonics noted in \cite{Banks} that become relevant whenever one starts to produce an axion with decay constant larger than $M_p$, and it leads us to suspect that the $\mathcal{O}(M_p)$ bound applies to not only an individual axion, but to the entirety of axion moduli space.  We provide further evidence for this conjecture with heuristic analyses of axions in type IIA and heterotic compactifications.

It must be emphasized that the axion moduli spaces considered are simulated to resemble true Calabi-Yau moduli spaces in a way that will be discussed more fully in \S 3--an unfortunately necessary simplification due to the fact that the relevant aspects of Calabi-Yau geometries are (slightly) beyond the realm of current mathematical understanding.  Nonetheless, if our analysis indeed applies beyond our simulated examples considered, this poses a significant challenge to axion inflation models that do not invoke monodromy \cite{Monodromy1,Monodromy2}.

In \cite{Swampland1, Swampland2}, it was pointed out that not all consistent low-energy effective theories admit consistent ultraviolet completions in string theory, and the sizeable collection of theories without an ultraviolet completion was deemed the `swampland' in contrast with the `landscape' of theories that do admit such a completion.  In \cite{wgc}, it was conjectured that any $U(1)$ gauge theory in the landscape must contain a state of mass $M$, charge $q$ satisfying the condition,
\begin{equation}
\frac{M}{q} \leq 1,
\end{equation}
in appropriate units.  This conjecture was primarily motivated by the absence of black hole remnants and has come to be known as the `weak gravity conjecture.'  It was further shown that the difficulty of finding a parametrically large axion decay constant emerges as a natural generalization of this weak gravity conjecture to $p$-form gauge fields.

In \cite{Cheung}, the weak gravity conjecture was extended to theories with multiple $U(1)$ gauge fields (and also found to present an interesting tension with the principle of naturalness).  We will show that the observed difficulty of finding a large axion moduli space diameter follows from this extension of the weak gravity conjecture, indicating that large axion moduli spaces are likely in the swampland of any consistent theory of quantum gravity.

The paper is structured as follows: in \S 2, we review axions in type IIB string compactifications and derive the bound (\ref{firstbound}) for Calabi-Yaus with simplicial K\"{a}hler cones.  We further discuss how this bound breaks down for non-simplicial K\"{a}hler cones.  In \S 3, we discuss and present the results of our moduli space simulation and statistical analysis.  In \S 4, we study axion moduli spaces in other corners of the string duality web.  In \S 5, we relate our bound to the weak gravity conjecture, and in \S 6, we present our conclusions.  A review of intersection theory and toric geometry is presented in the appendix, along with a simple example illustrating the relevant concepts for those unfamiliar with the subject.

\section{Axions in Type IIB Compactifications}

Before advancing to the geometric aspects of axions in type IIB compactifications, we present a brief review of the relevant aspects of type IIB compactifications.  A more detailed exposition can be found in \cite{BaumannMcAllisterBook}.

Compactification of type IIB string theory on a compact Calabi-Yau 3-fold results in a 4d $\mathcal{N} = 2$ supergravity theory.  Such Calabi-Yaus come equipped with a $(1,1)$-form $J$ known as the K\"ahler form, which takes values inside a strongly convex polyhedral cone whose interior is known as the K\"ahler cone.  In other words, given the set of generators $\omega_i \in H^{1,1}$ of the K\"ahler cone, the K\"ahler form may be written as $J = \omega_i t^i$ with coefficients $t^i >0$.  If the number of generators of the K\"ahler cone is equal to the dimensionality of the cone, the K\"ahler cone is said to be ``simplicial."  Otherwise, the number of generators outnumbers the dimensionality of the cone, and the K\"ahler cone is ``non-simplicial."    Figure \ref{nonsimplicial} in the appendix illustrates the difference between a simplicial cone and a non-simplicial one.

There is a cone within the vector space of 2-cycles of the Calabi-Yau known as the ``Mori cone," which is defined to be the dual cone of the K\"ahler cone.  In other words, the Mori cone is the set of all 2-cycles $C$ satisfying the condition $\int_C{\omega_i} \geq 0$ for all generators $\omega_i$ of the K\"ahler cone.  If the K\"ahler cone of the Calabi-Yau is simplicial, the Mori cone is also simplicial, and vice versa.

For the time being, we assume that the K\"{a}hler cone of our Calabi-Yau is simplicial, dealing with the more complicated case of a non-simplicial K\"{a}hler cone in the next subsection.  The R-R axion $C_0$ descends to an axion in $4d$, and pairs up with the dilaton $\Phi$ to form the complex axiodilaton,
\begin{equation}
\tau = C_0 + i e^{-\Phi}.
\end{equation}
Additional axions arise from integrating the NS-NS 2-form $B_2$, the R-R 2-form $C_2$, and the R-R 4-form $C_4$ over cycles of the appropriate dimensionality,
$$
b_i = \frac{1}{2 \pi\alpha'}\int_{\Sigma_{i}}{B_2}\,,~~~~c_i = \frac{1}{2 \pi\alpha'}\int_{\Sigma_{i}}{C_2}\,,\vartheta_i = \frac{1}{2 \pi (\alpha')^2}\int_{\Sigma_{i}}{C_4}.
$$
K\"{a}hler moduli similarly arise from integrating the K\"{a}hler form $J$ over the same 2-cycles,
$$
t_i = \frac{1}{\alpha'}\int_{\Sigma_{i}}{J}.
$$
The axions all carry a shift symmetry, which will be broken to a discrete subgroup by instanton effects, which from the string perspective is due to branes and worldsheet instantons wrapping the appropriate cycles \cite{Witteninstantons}.  In our normalization conventions, the periodicities of all of these axions is $2\pi$ (as opposed to the oft-used normalization in which the periodicity is $(2\pi)^2$).  The metric on axion moduli space is derived from the K\"{a}hler potential.  Working in large volume limit, we assume that the K\"{a}hler moduli are stabilized at $t_i > 1$ (in string units), so that non-perturbative corrections to the metric are exponentially suppressed and may be neglected.  The metrics (in 4d Planck units) on the moduli spaces of the axions $b_i$, $c_i$, and $d_i$ are then given respectively by,
\begin{equation}
b_i:\, G_{ij}^b = \frac{9}{4}\frac{N_{ik} t^k N_{jl} t^l}{(N_{mn}t^m t^n)^2 } -\frac{3}{2}\frac{N_{ij}}{N_{mn}t^m t^n } ,
\label{bmetric}
\end{equation}
\begin{equation}
c_i:\, G_{ij}^c = g_s^2 G_{ij}^b,
\label{cmetric}
\end{equation}
\begin{equation}
\vartheta_i:\, G_{ij}^{\vartheta} = g_s^2 \left(\frac{9}{2}\frac{t^i t^j}{(N_{mn} t^m t^n)^2} -3 \frac{(N^{-1})_{ij}}{N_{mn}t^m t^n }\right).
\label{varthetametric}
\end{equation}
Here, $N_{ij} = \kappa_{ijk}t^k$, and $\kappa_{ijk}$ are the triple intersection numbers
\begin{equation}
\kappa_{ijk} = \int_{Z}{\omega_i \wedge \omega_j \wedge \omega_k}\,,~~~~V(Z) = \frac{\alpha'^3}{6} \kappa_{ijk} t^i t^j t^k.
\label{volumeequation}
\end{equation}
in the chosen basis of $H^{1,1}(Z)$, $J = \alpha' \omega_i t^i$.

Clearly, in the region of theoretical control with $g_s < 1$, the size of the $c_i$ moduli space is suppressed relative to the size of the $b_i$ moduli space, so any bounds on the $b_i$ moduli space apply immediately to the $c_i$ moduli space.  We will focus primarily on the $b_i$ axions (and by extension, the $c_i$ axions) in this work, with more limited discussion of the $\vartheta_i$ axions.

To get a realistic model of nature featuring chiral matter, one must break half the supersymmetry to get an $\mathcal{N}=1$ supergravity theory by adding local objects such as D-branes.  Gravitational anomaly cancellation requires the positive tension of such D-branes to be cancelled by the presence of negative tension objects, most promisingly either $O3/O7$ orientifold planes or $O5/O9$ orientifold planes.  The orientifold action takes the form,
\begin{equation}
(-1)^{F_L} \Omega_{ws} \sigma,
\end{equation}
where $F_L$ is the number of fermions in the left-moving sector, $\Omega_{ws}$ reverses the orientation of the string worldsheet, and $\sigma$ is a geometric involution that flips the sign of the holomorphic 3-form $\Omega$ on the Calabi-Yau three-fold in the $O3/O7$ case.  Under this orientifold action, certain fields have even parity, and certain ones have odd parity, as shown in Table 3.2 of \cite{Grimmthesis}.  The cohomology group $H^{1,1}$ splits into a parity-even and a parity-odd part,
\begin{equation}
H^{1,1} = H_+^{1,1} \oplus H_-^{1,1}.
\end{equation}
If we select a basis $\{\omega_i\}$ for $H^{1,1}$ that decomposes into bases $\{\omega_i\}_+$ for $H_+^{1,1}$ and $\{\omega_i\}_-$ for $H_-^{1,1}$, we see that the $\{\omega_i\}_-$ will be projected out by the orientifold action, while the $\{\omega_i\}_+$ will be projected in.  Correspondingly in homology, if we select a basis of cycles $\{\Sigma_i\}$ for $H_2$ that decomposes as $\{C_i\}_+$ and $\{\Sigma_i\}_-$, the parity-odd $\{\Sigma_i\}_-$ will be projected out, while parity-even $\{\Sigma_i\}_+$ will be projected in.  One can maximize the number of moduli fields projected in by assuming that all $(1,1)$-forms are parity-even.

In both $O3/O7$ and $O5/O9$ compactifications, the $b_i$ axions have odd parity, so these will all be projected out upon taking $h^{1,1} = h_+^{1,1}$, $h^{1,1}_- = 0$.  However, the $c_i$ axions in $O5/O9$ compactifications and the $\vartheta_i$ axions in $O3/O7$ compactifications both have even parity, so these axions will be projected in, and our analysis goes through.

It was noticed in \cite{Nflation} that positivity of the $\kappa_{ijk}$ would prohibit a super-Planckian traversal of the inflaton.  We first make this argument rigorous, deriving a bound on the diameter of axion moduli space in the process.  We then explain why positivity of the $\kappa_{ijk}$ must indeed hold, as was argued in \cite{Grimm}.

To begin, we note that in the large volume limit, the metric on moduli space is independent of the axions.  Then, geodesics are represented simply by straight lines.  We define the diameter of the moduli space to be the length of the longest line contained in moduli space (without invoking monodromy), though for the purposes of inflation the relevant quantity is actually the length of the longest line ending at the origin, which is the minimum of the potential.  We label this quantity the radius, and for the $b_i$ moduli space it is given by
$$
r = \displaystyle\max_{x^i}\int_0^1 \sqrt{G_{ij}^b \frac{db^i}{ds} \frac{db^j}{ds}} ds\,,~~~~b^i =(1 - s) x^i
$$
\begin{equation}
=\displaystyle\max_{x^i}\sqrt{G_{ij}^b x^i x^j}
\end{equation}
We bound this quantity by bounding the integrand.  From the triangle inequality,
\begin{equation}
G_{ij}^b x^i x^j = \frac{9}{4}\frac{N_{ik} t^k x^i N_{jl} t^l x^j}{(N_{mn}t^m t^n)^2 } -\frac{3}{2}\frac{N_{ij} x^i x^j}{N_{mn}t^m t^n } \leq |\frac{9}{4}\frac{N_{ik} t^k x^i N_{jl} t^l x^j}{(N_{mn}t^m t^n)^2 }| + |\frac{3}{2}\frac{N_{ij} x^i x^j}{N_{mn}t^m t^n}|,
\label{triangleequation}
\end{equation}

If the $\kappa_{ijk}$ are all non-negative and the $t^i$ parametrize the volumes of the cycles, then all the terms in the above sums will be positive provided the $x^i$ are positive.  In such a case, we may bring the absolute values inside the sums without penalty, so that the bound becomes
\begin{equation}
G_{ij}^b x^i x^j  \leq \frac{9}{4}\frac{N_{ik} t^k |x^i| \cdot N_{jl} t^l |x^j|}{(N_{mn}t^m t^n)^2 } + \frac{3}{2}\frac{N_{ij} |x^i| |x^j|}{N_{mn}t^m t^n}.
\end{equation}
Without invoking monodromy, $|x_i| \leq  \pi$, and in the large volume limit, $t^i \gtrsim 1$.  Of course, the approximation in which non-perturbative corrections can be neglected gets better for increasing $t^i$, and t $t^i>1$ is a very generous lower bound.  Nonetheless, we shall adopt the condition $t^i >1$ for the sake of argument.  Hence, writing the above expression in a more suggestive way,
\begin{equation}
G_{ij}^b x^i x^j \leq \frac{9}{4}\frac{N_{ik} t^k |x^i|_{\mbox{max}}}{ N_{ik}t^k t^i_{\mbox{min}} } \frac{ N_{jl} t^l |x^j|_{\mbox{max}}}{N_{jl} t^l t^j_{\mbox{min}}}+ \frac{3}{2}\frac{N_{ij} |x^i|_{\mbox{max}} |x^j|_{\mbox{max}}}{N_{ij}t^i_{\mbox{min}} t^j_{\mbox{min}}}.
\end{equation}
we find, taking $|x^i|_{\mbox{max}} \rightarrow \pi$, $t^i_{\mbox{min}}\rightarrow 1$,
\begin{equation}
G_{ij}^b x^i x^j \leq \frac{15}{4} \pi^2,
\end{equation}
and hence,
\begin{equation}
r \leq \left( \frac{15}{4} \right)^{1/2} \pi.
\label{rbound}
\end{equation}
In particular, this bound is independent of not only the number of axions $N$, but even the geometry entirely--the triple intersections do not affect the bound.  If one cannot achieve a super-Planckian traversal with only one axion, one cannot do it with many.  If anything, increasing the number of species will make a super-Planckian traversal more difficult, as the many scalar fields will lead to a sizeable renormalization of the Planck mass \cite{dvali, Nflation}.  We further note that the triangle inequality used to justify (\ref{triangleequation}) is extremely conservative--in all examples studied, one may indeed keep the minus sign in the metric, so that,
\begin{equation}
r \leq \left( \frac{9}{4}  - \frac{3}{2} \right)^{1/2} \pi = \left( \frac{3}{4} \right)^{1/2} \pi.
\label{rbound2}
\end{equation}
By taking $x^i = \pi$ for all $i$, we find that this inequality is saturated.

To get inflation, one needs not only a super-Planckian radius, but also hierarchically light inflaton fields.  The masses of the axions wrapping the various cycles decay exponentially with the 2-cycle string-unit volumes as $e^{-t^i}$, so one must take $t^i > \lambda$ for some $\lambda \gg 1$ and some subset of $t^i$ to get a mass suitable for inflation.  This further shrinks the moduli space radius accessible during inflation.

A similar argument can be made for the size of $\vartheta_i$ moduli space, assuming that the $\kappa_{ijk}$ are large enough so that the denominators of both terms in the metric in (\ref{varthetametric}) will dominate their numerators.  This is indeed the case for projective-toric Calabi-Yau three-folds, and one expects the result to hold generally.

The preceding argument assumed positivity of the triple intersection numbers $\kappa_{ijk}$ and the K\"{a}hler moduli $t^i$.  We will now explain why positivity must indeed hold, elaborating on an observation made in \cite{Grimm}.

Of course, positivity of $\kappa_{ijk}$ is a basis-dependent statement.  If one goes to a new basis of $(1,1)$-forms, $\{\omega_i\}$, then the $\kappa_{ijk}$ will change as well.  The point is that there is a `good' basis in which the K\"{a}hler moduli correspond precisely to the volumes of the curves that generate the Mori cone.  Letting $C_i$ represent these curves, we choose our basis of $(1,1)$-forms to be the dual,
\begin{equation}
\frac{1}{\alpha'} \int_{C_i}{\omega_j} = \delta_j^i.
\end{equation}
The volumes of these curves are found by integration of the K\"{a}hler form, $J$, so that in this basis,
\begin{equation}
V(C_i) = \int_{C_i}{J}= \int_{C_i}{\omega_j t^j} = t^i \alpha'.
\end{equation}
Hence, in this 'good' basis, the string-unit volumes of the curves are simply the $t^i$.  The K\"{a}hler cone consists of all choices of K\"{a}hler form such that these volumes are positive, hence is given simply in this basis by $t^i > 0$.  Therefore, these $(1,1)$-forms $w_i$ generate the K\"{a}hler cone, and hence are Poincar\'{e} dual to nef divisors $D_i$.  The triple intersection of nef divisors is necessarily non-negative \cite{positivity}, so $\kappa_{ijk} \geq 0$ precisely when the K\"{a}hler moduli $t^i$ are positive (and represent the volumes of the appropriate holomorphic curves).  Hence, the positivity of the $\kappa_{ijk}$ and the $t^i$ assumed in the above argument has been established.  This rules out the possibility of large axion moduli spaces on Calabi-Yau three-folds with a simplicial K\"{a}hler cone, and brings into serious question the possibility of a cancellation in the denominators of the metrics of (\ref{bmetric}), (\ref{cmetric}), and (\ref{varthetametric}) that has long been recognized as a prerequisite for $N$-flation.  In particular, it places $h^{1,1}$$=$$2$ axion inflation models in the swampland of type IIB string theory (i.e. the set of consistent semiclassical theories not admitting a stringy completion), as all 2-dimensional cones are simplicial.

This argument also shows why the toy $N$-flation model proposed in \cite{Grimm} does not carry over to realistic Calabi-Yau compactifications.  This model relied on the negativity of intersection numbers of the resolved conifold with a cutoff imposed, read off from the prepotential of \cite{Hosono1, Hosono2, Candelas, GopakumarVafa}.  The resolved conifold is a non-compact space and hence not a complete algebraic variety, so the techniques of intersection theory utilized here do not apply to it.  Thus, although one can readily find negative intersection numbers for a non-compact Calabi-Yau like the resolved conifold, the story is very different for compact Calabi-Yaus.

\subsection{A Non-Simplicial K\"{a}hler Cone}

\begin{figure}
\begin{center}
\includegraphics[trim = 10mm 10mm 10mm 10mm, clip, width=7cm]{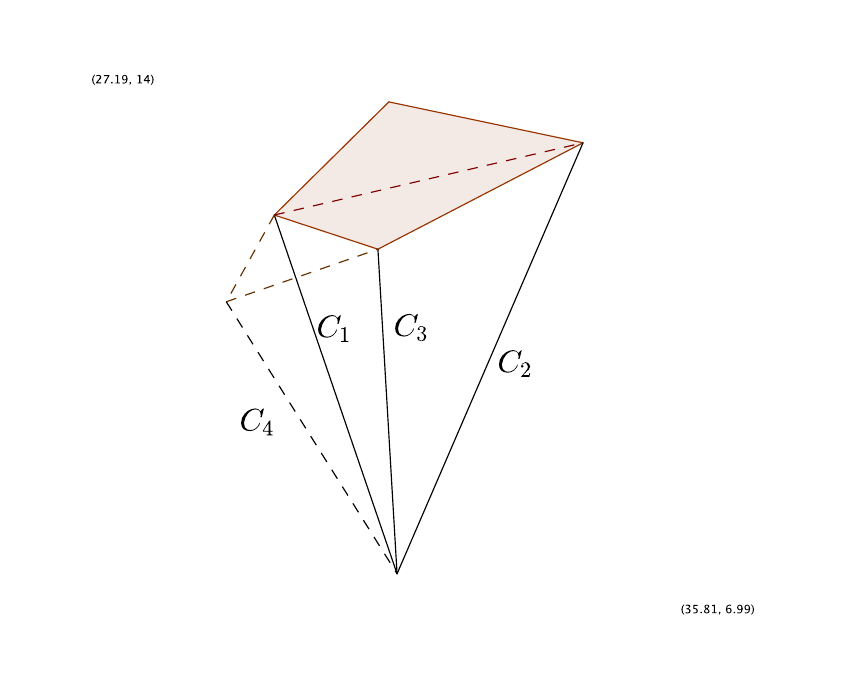}
\includegraphics[trim = 10mm 10mm 10mm 10mm, clip, width=7cm]{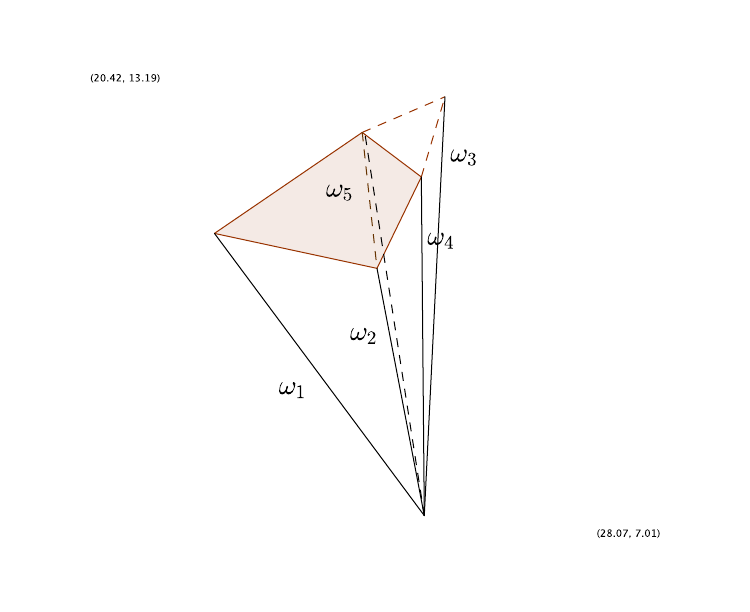}
\end{center}
\caption{A subset of a (non-simplicial) Mori cone (left) is dual to a superset of the (also non-simplicial) K\"{a}hler cone (right), and vice versa.}
\label{Morisuperset}
\end{figure}

We therefore turn to the case in which the K\"{a}hler cone of the Calabi-Yau is not simplicial, as depicted in Figure \ref{Morisuperset} (right).  This case will arise generically for Calabi-Yau hypersurfaces in toric varieties, so it cannot be ignored.  Recall that our bound in the simplicial case relied upon two conditions:
\begin{enumerate}
\item Positivity of the triple intersections $\kappa_{ijk}$.
\item $t^i > 1$ in string units, ensuring that the volumes of all curves are large so as to protect against worldsheet instanton and brane corrections to the metric and potential.
\end{enumerate}
In this case of a simplicial K\"{a}hler cone, as discussed, one may choose the basis of $(1,1)$-forms to correspond to the generators of the K\"{a}hler cone, which are dual to the generators of the Mori cone.  Clearly, then, the K\"{a}hler cone is given simply by the condition that all $t^i$ are positive, and each $t^i$ corresponds to the volume of a corresponding curve in the K\"{a}hler cone.  Non-negativity of intersections of nef divisors ensures non-negativity of $\kappa_{ijk}$.  Thus, conditions 1) and 2) are both satisfied.

In the non-simplicial case, we can no longer pick a basis that will cover the entire K\"{a}hler cone.  In Figure \ref{Morisuperset}, we must glue together two patches, e.g. one generated by $P_1=\{\omega_1,\omega_2,\omega_5\}$ and one generated by $P_2=\{\omega_2,\omega_4,\omega_5\}$.  Each of these patches is parametrized by $t^i \geq 0$, and each one will have $\kappa_{ijk} \geq 0$ because the generators are still nef.  Hence, condition 1) will hold on a patch-by-patch basis.  However, a subset of the K\"{a}hler cone is dual to a superset of the Mori cone, as shown in Figure \ref{Morisuperset}.  If we choose our patch of the K\"{a}hler cone to be generated by $P_1$, we will find there is no curve $C_4$ in the Mori cone for which $\int_{C_4}\omega_j = \delta_{j5}$.  Instead, the homology class $C_4$ dual to  $\omega_5$ lies outside the Mori cone.  Hence, we have lost the interpretation of the K\"{a}hler moduli as volumes of curves, and so condition 2) no longer holds.

Alternatively, we can find a basis of divisors $\{\omega_i\}, i \in \{1,...,h^{1,1}$, with $J = \omega_i t^i$ such that each $t^i$ corresponds to the volume of a generator of the Mori cone.  However, since the Mori cone is non-simplicial whenever the K\"{a}hler cone is simplicial, there are more than $h^{1,1}$ curves generating the Mori cone, so there will be additional generators of the Mori cone whose volumes are linear combinations of the $t^i$, Vol$(C) = \sum_i a_i t^i$.  Generically, some of the linear coefficients $\{a_i\},$ will be negative.  Defining $I_- = \{ i | a_i < 0 \}$ to be the set of indices with negative linear coefficient, we see that setting $t_k > 0$ for some $k \in I_-$ and $t_j = 0$ for all other $j$ would yield a negative volume for $C$.  As a result, we conclude that the divisor $D_j$ is not nef after all, so its triple intersections can be negative.

An alternative way to consider the two aforementioned bases is as follows: the first basis of divisors parametrizes only a subset of the K\"{a}hler cone, which is dual to a superset of the Mori cone.  At least one of the generators of this superset must therefore lie outside the Mori cone, so it does not correspond to an irreducible effective curve of the variety, and its K\"{a}hler modulus is no longer the volume of any such curve.  The second basis, on the other hand, parametrizes only a subset of the Mori cone, which is dual to a superset of the K\"{a}hler cone.  One of the generators of this superset must lie outside the boundary of the K\"{a}hler cone, so it is not nef, and its triple intersection numbers may be negative.

\begin{figure}
\begin{center}
\includegraphics[width=100mm]{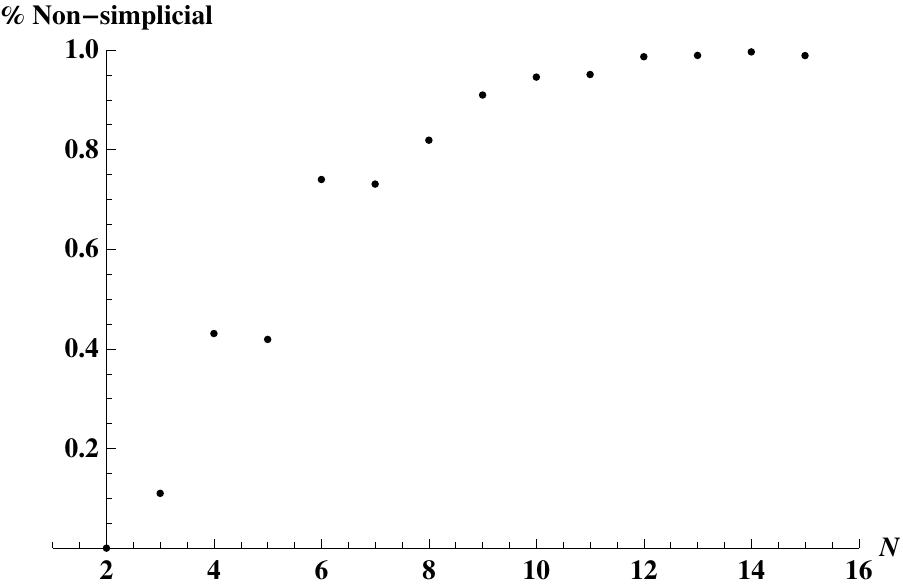}
\end{center}
\caption{The fraction of toric varieties with non-simplicial K\"{a}hler cone as a function of $N=h^{1,1}$  quickly approaches 1.}
\label{percentnonsimplicial}
\end{figure}

Thus, the na\"{i}ve argument forbidding a parametrically-large axion moduli space does not apply to a geometry with a non-simplicial K\"{a}hler cone.  In this case, either the positivity of the triple intersection numbers or the lower bound $t^i > 1$ must be sacrificed, depending on the basis of divisors with which one chooses to work.  In other words, there is no longer a 'good' basis for which both the $t^i$ correspond to volumes of curves and $\kappa_{ijk} \geq 0$.  This is not just a pathological exception, but rather an arbitrarily-selected algebraic variety will almost always have a non-simplicial K\"{a}hler cone, as shown in Figure \ref{percentnonsimplicial}.  At the time of publication of \cite{KatzCox}, all known K\"{a}hler cones of Calabi-Yau three-folds were simplicial, as the GKZ chambers of several ambient toric varieties related by sequences of trivial flops were found to patch together to form a simplex.  However, it was later shown in \cite{CoxKatzCounterexample} that the K\"{a}hler cone of a Calabi-Yau is not necessarily contained in the union of the K\"{a}hler cones of varieties related by sequences of trivial flops, so simpliciality of the latter space need not imply simpliciality of the former.  Furthermore, we find in our analysis that there exist Calabi-Yaus embedded in toric varities with no trivial flops (i.e. no paper walls at the boundary of the nef cone, in the language of our appendix) with non-simplicial K\"{a}hler cone--a simple counterexample to the speculation that the union of all K\"{a}hler cones of ambient varieties related by trivial flops should form a simplex.  We expect more generally that most Calabi-Yaus will have non-simplicial K\"{a}hler cone, and so we expect that the bound derived in (\ref{rbound}) therefore will not apply to the majority of Calabi-Yau compactifications.

However, not only is the metric on the moduli space of the $b_i$ and $\vartheta_i$ axions altered in the non-simplicial case, but so too is the range over which the axions can vary.  In a generic $\mathcal{N}=1$ compactification, this range is dependent upon which effective cycles give rise to superpotential terms when they are wrapped by Euclidean branes or worldsheet instantons.  Whether or not a cycle contributes to the potential is further dependent upon homological data of the cycle and flux data of the compactification \cite{Halverson}.  For our purposes, we define the ``diameter" of axion moduli space to be the maximal distance between two points in the axion moduli space whose boundaries are given by assuming that all generating curves of the Mori cone introduce terms to the potential.  Relaxing this assumption could in some cases improve the prospects for axion inflation via decay constant alignment \cite{KNP}, but we will leave a study of such alignment models for future work.\\ \indent
In the basis in which the K\"{a}hler moduli $\{t^i\}$ represent volumes of generating curves $\{C_i\}$ of the Mori cone, branes and worldsheet instantons will give contributions to the potential of the form,
\begin{equation}
(\Lambda_i^b)^4 \left( 1-\cos{b_i} \right)\,,~~~~ (\Lambda_i^c)^4 \left( 1-\cos{c_i} \right)
\end{equation}
where $b_i$ and $c_i$ come from integrating the NS-NS and R-R 2-forms, respectively, over the curve $C_i$, and $\Lambda_i^b$ and $\Lambda_i^c$ are dynamically-generated scales.  From the low-energy perspective, these scales are proportional to the instanton action, which translates to the string perspective as $\Lambda^4 \sim e^{-t}$, where $t$ is the volume of the 2-cycle in question.  Similarly, there will be contributions to the potential of the form,
\begin{equation}
(\Lambda_i^\vartheta)^4 \left(1-\cos{\vartheta_i}\right),
\end{equation}
for the axions $\vartheta_i$ coming from integration of $C_4$ over the generators of the cone of effective 4-cycles, with $\Lambda^4 \sim e^{-\tau}$ for a $4$-cycle of volume $\tau$.  Said contributions appear regardless of whether or not the K\"{a}hler cone is simplicial.  However, in the non-simplicial case, there will be additional cycles, corresponding to the extra generators of the Mori cone.  A generator $C_*$ with volume given by $\sum_i a_i t^i$ for specified coefficients $a_i$ will yield a contribution to the potential of the form,
\begin{equation}
(\Lambda_{*}^b)^4 \left(1-\cos{\sum_i{a_i b_i}}\right)
\label{lambdastar}
\end{equation}
and similarly for the $\{c_i\}$ and $\{\vartheta_i\}$ axions.  We see that the concept of the fundamental domain of axion moduli space, $b_i \in [-\pi, \pi]$, is basis-dependent and therefore ill-defined in the absence of a unique 'good basis.'  If we let a different set of generating curves serve as our basis, then the boundaries of moduli space will change as well.  The distance over which the inflaton fields will vary during the course of an slow-roll inflationary period must be basis-independent, and so it is worth considering the distance from the maxima of the inflationary potential to their nearby minima.

At first, one might think that the extra contributions to the potential within the fundamental region $b_i \in [-\pi, \pi]$ coming from the extra cyles in the non-simplicial case would necessarily restrict the range of the inflaton fields, since there will be extra contributions to the potential in the fundamental region that are not present in the simplicial case, which may spoil slow-roll.  We will see that this is indeed the case, but only in the right basis.  Suppose, on the other hand, that the scales $\Lambda_i$ of the curves whose volumes are parametrized by the $t^i$ are much smaller than the scales $\Lambda_*$ whose volumes are parametrized by linear combinations $\sum_i a_i t^i$.  Then, the potential contributions from the former set of curves will be negligible compared to the latter, and the effective range accesible to the axions during slow-roll may have nothing to do with the cube $b_i \in [-\pi,\pi]$.

To get around this difficulty, we can choose our basis of $(1,1)$-forms $\omega_i$ so that the (linearly independent) Mori generators of smallest possible volume are the ones whose volumes are parametrized by the $t^i$.  Since $\Lambda^4 \sim e^{-t^i}$, the larger potential contributions will come from the smaller cycles, and so in this basis they will form a cube $b_i \in [-\pi,\pi]$.  Other cycles whose volumes $\sum_i a_i t^i$ are not significantly greater will introduce significant contributions within this cube, essentially imposing the restriction $\sum_i a_i b_i \in [-\pi,\pi]$.  This is similar to the observation of \cite{Banks} that taking $t^i < 1$ will introduce further harmonics to the potential, i.e. $\Lambda_i^4 (1-\cos{n \phi})$, which will restrict the range to $b_i \in [-\pi/n,\pi/n]$.

Often, there are additional irreducible curves in the interior of the Mori cone, but the volumes of these curves will be positive linear combinations with $\mathcal{O}(1)$ coefficients of the volumes of the Mori generators and so will typically be larger than the generating curves.  Large volumes imply subdominant contributions to the potential, so we neglect these curves in our analysis.

It is therefore possible, at least in principle, to select a basis in which the effective field range accessible for inflation is not enhanced (i.e. is approximately a cube $b_i \in [-\pi, \pi]$) and the K\"{a}hler moduli correspond to volumes of curves generating the Mori cone.  If the K\"{a}hler cone of the Calabi-Yau is simplicial, then the volumes of all of the generating curves of the Mori cone are represented by K\"{a}hler parameters, the large volume region of parameter space is given simply by the condition $t^i \gtrsim 1$ for all $i$, the triple intersection numbers are all non-negative, and the dominant cosine contributions to the axion potential all have a single axion in their argument (i.e. the mass matrix is diagonal).  In the non-simplicial case, there will be some Mori cone generators whose volumes are given by linear combinations of the K\"{a}hler moduli, so the large volume region of parameter space is given by $t^i \gtrsim 1$ along with additional linear restrictions on the $t^i$.  The triple intersection numbers can be negative because the basis divisors are not necessarily nef, and there may be additional dominant cosine contributions to the axion potential whose arguments are linear combinations of axion fields (i.e. the mass matrix is not diagonal).  This basis is the unique 'good basis' in the simplicial case, and it is the basis in which the bound of (\ref{rbound}) is easily derived.  In the non-simplicial case, it provides the clearest window for understanding and analysis of the diameter of axion moduli space, though no similar bound can be derived given the negativity of the triple intersection numbers.  Instead, one is led naturally to a statistical approach, measuring the diameter of moduli space in a sampling of Calabi-Yau manifolds, and comparing the shrinking of the metric on axion moduli space with the Pythagorean gain from increasing the number of axions.  In the following section, we describe our analysis and present the most interesting results.

\section{Statistical Analysis and Results}

As stated, our analysis considers a large collection of 'simulated' Calabi-Yau axion moduli spaces, rather than true Calabi-Yau axion moduli spaces.  This is due to the fact that it is currently unknown how to compute the entire K\"{a}hler cone of a generic Calabi-Yau--even where we expect the cone to be convex rational polyhedral (see the appendix for further discussion).

We want a way to simulate Calabi-Yau axion moduli spaces by first simulating K\"{a}hler cones of Calabi-Yaus.  There are several conditions that these must satisfy, coming from both a physical and a mathematical perspective.  From the physics point of view, they must have positive definite metrics (\ref{bmetric}), (\ref{cmetric}), (\ref{varthetametric}) on the entire K\"{a}hler cone (i.e. whenever the volumes of the generating curves of the Mori cone $\sum_i a_i t^i$ are greater than 0).  They must also have positive Calabi-Yau volume (\ref{volumeequation}) over the entire K\"{a}hler cone.  From the mathematics point of view, triple intersection numbers $\kappa_{ijk} = \int{\omega_i \wedge \omega_j \wedge \omega_k}$ must be non-negative whenever $\omega_i$, $\omega_j$, and $\omega_k$ are nef.  In particular, this means that all intersection numbers must be non-negative whenever the K\"{a}hler cone is simplicial.

The physical constraints impose strong restrictions on the K\"{a}hler cone.  In the two-modulus case, enforcing positive-definiteness of the metric whenever $t^1, t^2 > 0$ suffices to enforce non-negativity of all triple intersection numbers--even without the mathematical constraint coming from nefness of the divisors.  Considering the generalization to a higher-dimensional Calabi-Yau with simplicial K\"{a}hler cone and so taking $t^k \rightarrow 0$ for all $k \neq i, j$, one finds the restriction $\kappa_{iii}, \, \kappa_{iij} \geq 0$ for $i \neq j$.  (Positive definiteness of the metric may also suffice to bound $\kappa_{ijk} \geq 0$, $i \neq j \neq k$ in the simplicial case, but it is much more difficult to show.)  Furthermore, positivity of the $\kappa_{ijk}$ is nowhere near sufficient to guarantee positive definiteness of the metric on the entire K\"{a}hler cone.  A randomly generated set of $\kappa_{ijk}$ and a randomly generated set of K\"{a}hler cone constraints $\sum_i a_i t^i > 0$ will almost certainly fail to produce a positive definite metric on all of moduli space.

It is very difficult, therefore, to simulate a Calabi-Yau moduli space from scratch.  However, there is a way to produce moduli spaces satisfying the constraints mentioned above: instead of computing the K\"{a}hler cone of the entire Calabi-Yau, use instead the K\"{a}hler cone of the ambient toric variety, with triple intersection numbers given by the triple intersection numbers of divisors in the embedded Calabi-Yau.  These cones satisfy all the prerequisite physical and mathematical constraints, and so provide the best available tool for statistical analysis of true Calabi-Yau moduli spaces.  These objects were used, for instance, in the related statistical analysis of \cite{HeavyTails}, and they define what precisely we mean by the term, 'simulated moduli spaces.'

In \cite{KreuzerSkarke}, M. Kreuzer and H. Skarke famously classified all 473,800,776 reflexive polyhedra in 4d which yield Calabi-Yau three-fold hypersurfaces.  The complete list can be found at \cite{PALP}.  Here, we present the results of our analysis of 8395 simulated axion moduli space radii for toric Calabi-Yau threefolds in the Kreuzer-Skarke dataset.

To begin, we arbitrarily select a collection of reflexive, favorable polytopes for Hodge numbers $h^{1,1} = N = 2,3,4,...15$.  We then compute triangulations of these polytopes using the triangulation algorithm in Appendix B of \cite{HeavyTails}, saving a handful (up to 16) regular, star, complete triangulations for each polytope, using Sage as an interface.  For each such triangulation, we compute the triple intersection numbers $\kappa_{ijk}$ between each of the divisors, along with the relevant Mori cone data, which allows us to determine the GKZ chamber for the triangulation in question.

For this GKZ chamber, corresponding to the nef cone of the ambient toric variety associated to the given triangulation of the polytope, we move to a 'good basis' of divisors discussed previously.  We insist that all curves of the ambient toric variety should have volume $>1$, so that we stay in the red regions in Figure \ref{LVSpicture}.  Constrained to lie within this realm, we minimize the volume $V=\frac{1}{6} \kappa_{ijk} t^i t^j t^k$ appearing in the denominators of the metric on moduli space.

We look at how the largest eigenvalue of the metric varies with $N$.  The crucial assumption in the literature to this point has been that a cancellation can occur in the overall volume of the Calabi-Yau, so that the largest eigenvalue of the metric stays approximately constant while the length of the diagonal grows as $\sqrt{N}$.  We compute a rough estimate of the radius allowed in the large volume region of parameter space via,
\begin{equation}
\tilde{f} = \pi \sqrt{N} f,
\end{equation}
where $f$ is the square root of the maximum eigenvalue of the metric in the large volume regime of Nef($X$).  As realized in \cite{kineticalignment}, this gives a good idea of the physical radius of the cube $b_i \in [-\pi,\pi]$, since the largest eigenvalue of the metric almost always points along a diagonal of the cube.  This happens increasingly more often as $N$ increases.

\begin{figure}
\begin{center}
\includegraphics[width=100mm]{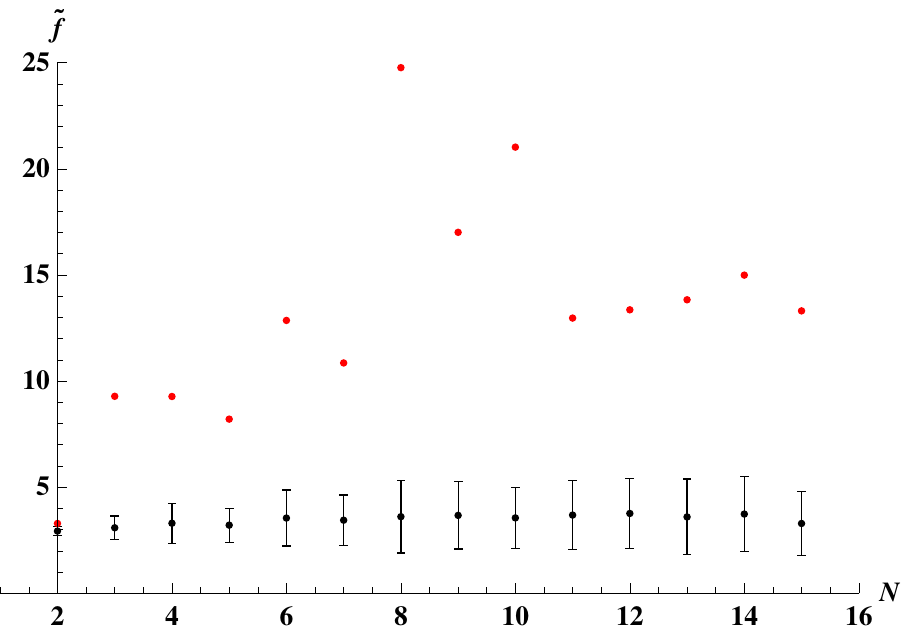}
\end{center}
\caption{The average value of $\tilde{f}  = \pi \sqrt{N} f$ is shown in black, with error bars indicating the standard deviation over the sample of Calabi-Yaus studied.  Here, $f$ is the square root of the largest eigenvalue of the metric $G_{ij}^b$.  The maximum $\tilde{f}$ at each $N$ is shown in red.  The average does not vary greatly with $N$.}
\label{baxionplot}
\end{figure}

\begin{figure}
\begin{center}
\includegraphics[width=100mm]{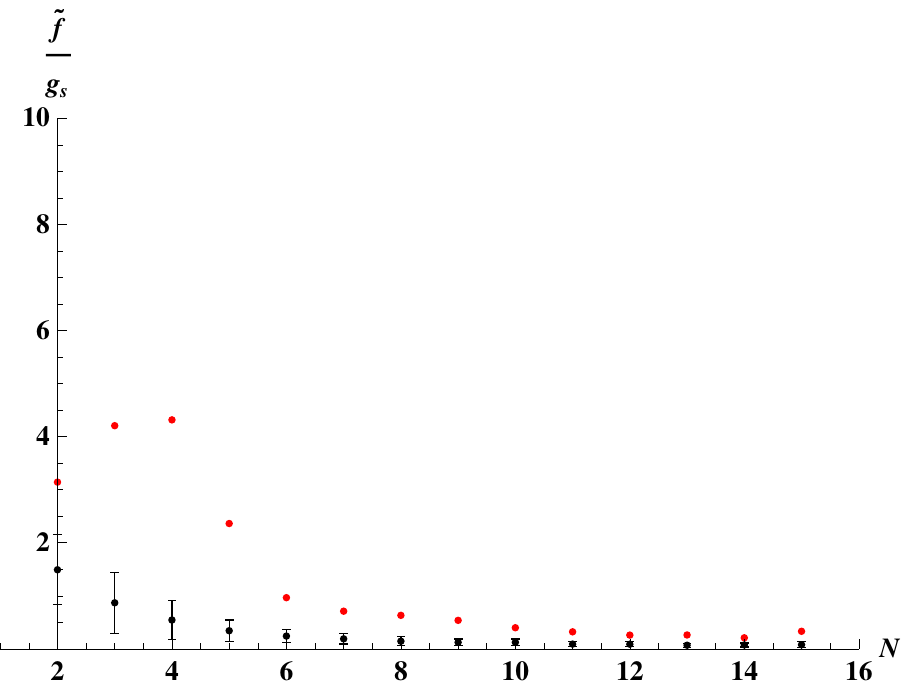}
\end{center}
\caption{The average value of $\tilde{f} / g_s = \pi \sqrt{N} f / g_s$, where $f$ is the square root of the largest eigenvalue of the metric $G_{ij}^\vartheta$, is shown in black, with error bars indicating the standard deviation over the sample of Calabi-Yaus studied.  The maximum $\tilde{f}$ at each $N$ is shown in red.  The average decreases with increasing $N$.}
\label{thetaaxionplot}
\end{figure}

The results of this analysis are shown in Figures \ref{baxionplot} and \ref{thetaaxionplot} for the $b_i$ axions and the $\vartheta_i$ axions, respectively.  The black dots indicate the average maximum $\tilde{f}$ within the aforementioned region of parameter space, the error bars indicate the spread (standard deviation) of this quantity, and the red dots indicate the maximum values observed for each $N$.  Concentrating first on the $b_i$ axions in Figure \ref{baxionplot}, we note first off that while the bound (\ref{rbound2}) on the radii of axion moduli space with a simplicial K\"{a}hler cone restricts $\tilde{f}$ severely for $N=2$, it breaks down for non-simplicial K\"{a}hler cones, and indeed for $N \geq 3$ we see significantly larger $\tilde{f}$.  The maximum $\tilde{f}$ at each $N$ does appear to grow slightly with $N$, but this growth is significantly smaller than the large variations with Calabi-Yaus of a given Hodge number, which are indicative of the heavy-tailedness of the eigenvalue distribution observed in \cite{HeavyTails}.  However, it is clear that the average $\tilde{f}$ is almost perfectly constant for increasing $N$.  We find no evidence for parametric enhancement of the moduli space diameter.

The situation gets even worse when it comes to the $\vartheta_i$ axions.  Recall first that the metric on $\vartheta_i$ moduli space is suppressed by $g_s^2$, so that the y-axis of Figure \ref{thetaaxionplot} is $\tilde{f} / g_s$, rather than $\tilde{f}$.  Avoiding large radiative corrections to the Planck mass from string loops requires $g_s^2 \lesssim 6\pi/N$ \cite{Nflation}, which means that $\tilde{f}$ must be scaled down by yet another factor of $\sqrt{N}$ for very large $N$, in addition to the loss depicted in Figure \ref{thetaaxionplot}, and even bearing in mind that $\tilde{f}$ has a free factor of $\sqrt{N}$ included in its definition.  For $\vartheta_i$ axions, therefore, it appears that the prospects of acheiving a super-Planckian displacement are greatly diminished as $N$ gets large, provided one works in the large volume region $t^i > 1$.

These findings are easy to understand from the axion moduli space metrics.  On $b_i$ moduli space, the line element is the difference of two terms,
\begin{equation}
ds_b^2 = \frac{9}{4}\frac{\kappa_{ikl} t^l t^k db^i \kappa_{j mn} t^l t^m db^j}{(\kappa_{lmn} t^l t^m t^n)^2 } -\frac{3}{2} \frac{\kappa_{ijk} t^k db^i db^j}{\kappa_{lmn} t^l t^m t^n } .
\end{equation}
Since each of the sums implied by the Einstein summation convention is over $N$ elements, we expect the line element to scale as,
\begin{equation}
ds_b^2 \sim \frac{\mathcal{O}(N^6)}{\mathcal{O}(N^6)} - \frac{\mathcal{O}(N^3)}{\mathcal{O}(N^3)}= \mathcal{O}(N^{0}).
\end{equation}
The ranges of all of the axions are independent of $N$, so we expect the radius to scale as $\mathcal{O}(N^0)$ i.e. to be independent of $N$.  For the $\vartheta_i$ axions,
\begin{equation}
ds_{\vartheta}^2= g_s^2 \left(\frac{9}{2}\frac{t^i t^j d\vartheta_i d\vartheta_j}{(\kappa_{lmn}t^l t^m t^n)^2} -3 \frac{(\kappa_l^{\,ij} t^l)^{-1} d\vartheta_i d\vartheta_j}{\kappa_{lmn} t^l t^m t^n }\right).
\label{thetalineelement}
\end{equation}
Hence we expect scaling as,
\begin{equation}
ds_\vartheta^2 \sim \frac{\mathcal{O}(N^2)}{\mathcal{O}(N^6)} - \frac{\mathcal{O}(N^{-1})}{\mathcal{O}(N^3)}= \mathcal{O}(N^{-4}).
\end{equation}
We therefore expect the radius of $\vartheta_i$ moduli space to shrink with $N$, as observed.

Of course, as we have seen, provided the K\"{a}hler cone of the Calabi-Yau is non-simplicial, one expects to find some negative triple intersection numbers $\kappa_{ijk}$ while working in the 'good basis.'  Thus, one expects some cancellation between the terms in the denominator of the line elements.  However, there is no good reason to think that this cancellation in the denominator should be so perfect as to allow one to get a volume that is independent of $N$ for a typical Calabi-Yau, allowing axion decay constants that do not vary at all with $N$.  More likely, the cancellation will be partial and $N$-dependent, so that enhancement of the radius will be partial and much less than the Pythagorean $\mathcal{O}(N^{1/2})$ hoped for by $N$-flation.  What's more, if one could achieve such near-perfect cancellation as some point in K\"{a}hler moduli space, it is likely that one could find a nearby point in K\"{a}hler moduli space in which the negative terms overwhelm the positive ones, resulting in a unphysical negative volume in some region of moduli space.  Acheiving a near-perfect cancellation of positive contributions to the volume by negative ones without permitting negative contributions to become dominant at any point of moduli space would require a substantial fine-tuning of the geometry.

By this analysis, we therefore expect to find that the radius of $b_i$ moduli space should not grow with $N$, but that suitable cancellations should occur in the denominator of the metric in the non-simplicial case and boost $\tilde{f}$ above its bounded value $\pi \sqrt{\frac{3}{4}}$, by an amount that is largely independent of $N$.  We further expect that such cancellations should also occur in $\vartheta_i$ axion moduli space, but that the inverse scaling of the radius with $N$ will dominate these cancellations, so that the peak of the $\vartheta_i$ moduli space radius is at the smallest values of $N$ that admit non-simplicial K\"{a}hler cones.  This is precisely what is observed in the eigenvalues of the moduli space metric metric in Figures \ref{baxionplot} and \ref{thetaaxionplot}, and even explains the peak in the $\vartheta_i$ metric eigenvalues at $N=3,4$.

Looking at the metric for the $\vartheta_i$ axions, the following idea for acheiving a large radius may occur: arrange the volume of the Calabi-Yau to be independent of some $\vartheta_a$ (so that $\kappa_{ija} = 0$ for all $i,j$), and make the first term of (\ref{thetalineelement}) arbitrarily large in the $a$ direction by taking $t^a \rightarrow \infty$.  Unfortunately, this idea fails, as it would result in a non-invertible matrix $N_{ij}$, so that the metric would become singular.  However, it is not so difficult to find examples for which $\kappa_{aaa}=0$: if one could arrange for the other cycles $t^i \neq t^a$ to be significantly less than 1, then the line element in the $a$ direction would scale approximately as,
\begin{equation}
ds_\vartheta^2 \approx \frac{(t^a d\vartheta_a)^2}{(t^i)^2 (t^a)^4},
\end{equation}
which would get arbitrarily large for $t^i$ arbitrarily small.  Once again, though, such a move is not allowed, as instanton corrections become important for $t^i \rightarrow 0$ and spoil our attempted solution.

At this point, the idea of $\mathcal{O}(N^{1/2})$ scaling of axion moduli space has already been called into question.  In most of our simulated axion moduli spaces, there is no enhancement of metric eigenvalues above the bound (\ref{rbound2}).  In the rare cases in which such enhancement occurs, it does not depend noticeably on $N$, in agreement with our arguments in the preceding paragraphs.  Rather, a more important criterion than dimensionality of the axion moduli space is non-simpliciality of the K\"{a}hler cone--without this, it is impossible to get any enhancement whatsoever.

However, we have neglected up until this point the additional contributions that arise in compactifications on a non-simplicial K\"{a}hler cone, discussed in \S 2.  The potential contributions coming from branes/worldsheet instantons wrapping a cycle become large whenever the volume of that cycle becomes small, scaling as $e^{-t^i}$.  To determine the effective range of the axions, we therefore impose the following algorithm to determine which curves $S := \{C_{i_1},...C_{i_n}\}$ will yield significant contributions to the potential.  We perform this analysis only for the $b_i$ axions (and by extension, the $c_i$ axions), as a similar analysis for the $\vartheta_i$ axions would require more intimate knowledge of the cone of effective divisors, which is not well understood.
\begin{enumerate}
\item Consider the $M \geq N$ curves generating the Mori cone, $C_1,...C_M$, and sort them by volumes vol($C_j$)$=\sum_i a_i^j t^i$, $j =1,...,M$.
\item Process each curve in order of volume: if the charge vector $a_i^j$ of this curve is linearly independent from the curves already in the set $S$, add it to $S$.  At the end of this step, $S$ will contain precisely $N$ elements, since it will contain a maximal set of linearly independent vectors of an $N$-dimensional vector space.  In the aforementioned `good basis' of $(1,1)$-forms for a non-simplicial K\"{a}hler cone, these will simply satisfy $a_i^j = \delta_i^j$.
\item Next, look at the curves not yet in $S$.  For each such curve $C_k$, consider the subset of curves in $S$ whose volumes are all at least 1 (in string units, as always) larger than this curve, $C_k$.  If $a_i^k$ is contained in the span of the vectors $a_i^j$ of this subset of curves, then $C_k$'s contribution to the potential is considered subdominant, and it is not added to $S$.  If, on the other hand, $C_k$ is not contained in the span of these curves, then it is assumed that $C_k$ will contribute significantly to the potential, and $C_k$ is included in $S$.  This step ensures that we are only considering potential contributions that are not minuscule compared to the dominant potential contributions in in each direction of moduli space, and is automatically skipped in the simplicial case where there are no significant additional contributions in the large volume limit.
\item The effective range of the axions $b_i$ is taken to be $\sum_i a_i^j b_i \in [-\pi, \pi]$ for all $j$ such that $C_j \in S$.  This enforces a cutoff on the effective range of the axions wherever there is a large potential contribution.
\end{enumerate}

\begin{figure}
\begin{center}
\includegraphics[width=100mm]{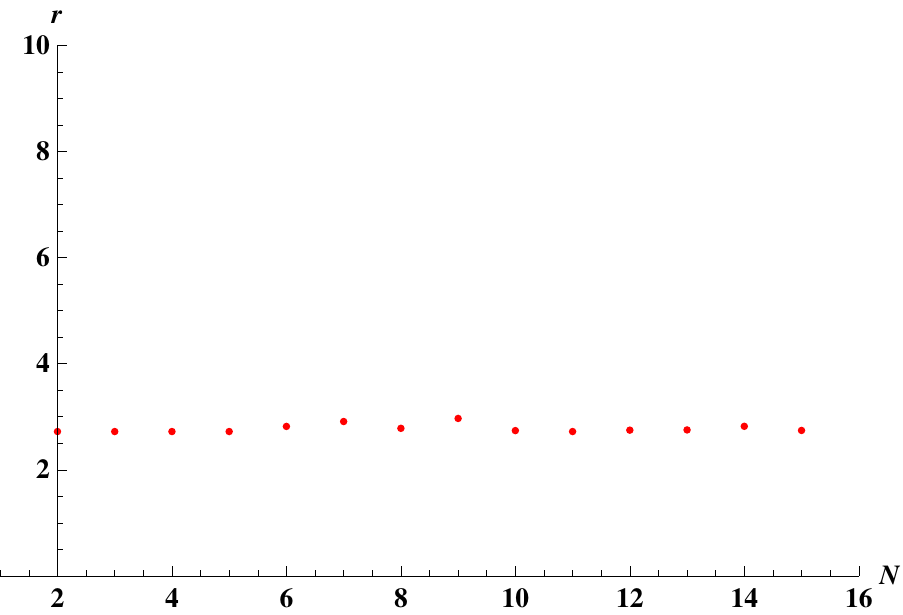}
\end{center}
\caption{The largest effective radius of axion moduli space for each $N$ is shown.  Once we account for additional contributions to the potential arising in compactifications on Calabi-Yaus with non-simplicial K\"{a}hler cone, the effective range accessible to the axions is cut off (compare with Figure \ref{baxionplot}).  In all simulated axion moduli spaces considered, the effective radius is cut down to less than $\pi \, M_p$, well within the bound (\ref{rbound}) derived in the simplicial case.}
\label{keyplot}
\end{figure}

The results of this algorithm and subsequent analysis are shown in Figure \ref{keyplot}.  Comparison with Figure \ref{baxionplot} tells a remarkable story: in \emph{all} cases in which some eigenvalue of the metric becomes large (necessarily a non-simplicial case), contributions from the additional curves generating the Mori cone become important, and cut off the eigendirection of largest eigenvalue, shrinking the radius of moduli space accessible to the axions to less than $\pi \, M_p$, well within the derived simplicial bound (\ref{rbound}), and roughly equal to the observed simplicial bound $(\ref{rbound2})$.

This is rather reminiscent of the situation described in \cite{Banks} for a single axion--although it is possible to find axions with parametrically large decay constant, these axions contain appreciable harmonics that cut off their effective range.  As soon as one thinks he or she has found a way to stretch out the axion potential beyond the Planck scale, new contributions start popping up to prohibit it.  The results in Figure \ref{keyplot} suggest that this happens in the multi-axion case just as well as the single axion case, prohibiting a large axion traversal no matter how one attempts to circumvent the issue.

\section{Axions in Other Corners of the Duality Web}

So far, we have seen evidence from type IIB compactifications that the $\mathcal{O}(M_p)$ bound on individual axion decay constants should generalize to the radius of the entire axion moduli space, rather than applying to each axion individually.  In this section, we provide arguments from compactifications in type IIA and heterotic string theory to support this conclusion.  Mathematical difficulties make the detailed analysis we have performed for type IIB axions rather difficult to achieve, so we content ourselves with a much more heuristic approach.  The details of axion moduli spaces in the following subsections come from the classic paper \cite{WittenAxions}.

\subsection{Axions in Heterotic Compactifications}

As in type IIB compactifications, there is a model-independent axion that arises in heterotic string compactifications on a Calabi-Yau $Z$ from dualizing the 4d component of the 2-form $B_{\mu\nu}$.  However, we are concerned with the behavior of axion moduli spaces as the number of axions $N$ grows large, as the single axion case was addressed in \cite{Banks}.  Thus, we focus on the model-dependent axions arising from integrating the 2-form $B_2$ over the curves of the manifold.  Given a basis of $(1,1)$-forms $\omega_i$ satisfying $\int_{C_j}\omega_i = \delta_{ij}$ for curves $C_j$, one finds a metric on axion moduli space,
\begin{equation}
g_{ij}^H=\frac{\pi}{ g_s^2 l_s^4}\int_{Z}{\omega_I \wedge * \omega_j}.
\label{hetmetric}
\end{equation}
But, using the Calabi-Yau identity,
\begin{equation}
* \omega_i = - J \wedge \omega_i + \frac{J \wedge J}{4 V} \int_{Z}{\omega_i \wedge J \wedge J}
\end{equation}
for K\"{a}hler form $J$, along with the identities $V = \frac{1}{6} \int_Z{J \wedge J \wedge J}$ and $M_p^2 = 4 \pi \frac{V}{g_s^2 l_s^8}$, (\ref{hetmetric}) simplifies to,
\begin{equation}
g_{ij}^H/M_p^2 = \frac{9}{4}\frac{\kappa_{ikl}t^k t^l \kappa_{jmn} t^m t^n}{(\kappa_{ijk}t^i t^j t^k)^2} - \frac{3}{2}\frac{\kappa_{ijk} t^k}{\kappa_{ijk}t^i t^j t^k},
\end{equation}
which is the same as the metric (\ref{bmetric}).  The analysis is identical, therefore, to the type IIB case.

\subsection{Axions in Type IIA Compactifications}

Axions arise in Type IIA from integrating the NS-NS 2-form $B_2$ over $2$-cycles and from integrating the R-R 3-form $C_3$ over $3$-cycles.  The former gives rise to the same situation studied in the type IIB case.  The latter gives rise to a moduli space metric,
\begin{equation}
g_{ij}^{IIA} = \frac{\pi}{l_s^2}\int_Z{\omega_i \wedge * \omega_j},
\end{equation}
where $\omega_i$ are representatives of $H^2(Z)$.  We assume that it is possible to take our basis of $3$-cycles $C_i$ to be calibrated with respect to the holomorphic 3-form $\Omega$, so that their volumes can be expressed as vol($C_i$)$=\int_{C_i}\Omega$.  Thus, $\Omega = \sum_i \omega_i t^i$, with $\int_{C_j}\omega_i = \delta_{ij}$.  We further assume that these $\omega_i$ generate some cone in a real slice of complex structure moduli space from which the axions arise by complexification (analogous to the K\"{a}hler cone for K\"{a}hler moduli space, and presumably related to it by mirror symmetry), and that the calibrated cycles $C_i$ generate some dual cone of $3$-cycles (analogous to the Mori cone).  Without these assumptions, it is not clear how to express the volumes of the $3$-cycles in terms of the $t^i$ or how to estimate potential contributions from branes wrapping the cycles.  Under these assumptions, the metric becomes,
\begin{equation}
g_{ij}^{IIA} = \frac{g_s^2 M_p^2}{4}\frac{\int_Z \omega_i \wedge * \omega_j}{(\sum_i \omega_i t^i) \wedge * (\sum_j \omega_j t^j)}.
\label{Ametric}
\end{equation}
As with the axions coming from $C_4$ integrated over $4$-cycles in type IIB, detailed analysis would require more information about the analog of the Mori cone for $3$-cycles, which is not well understood.  Instead, we simply estimate the scaling of the moduli space radius with the number of axions $N$.  Since the denominator consists of two sums of $N$ elements, we expect (\ref{Ametric}) to scale as $1/N^2$.  Once again, this will cancel the gain from traveling along the diagonal, and so we na\"ively expect $\mathcal{O}(N^0)$ scaling of the radius with $N$, as has been observed in previous cases.

The problem in all situations boils down to this: as one increases $N$, the volume of the Calabi-Yau increases as well.  Since the 4d Planck mass scales as $V^{1/2}$, the Planck-unit metric will scale down with $N$, even as the dimensionality of the metric scales up with $N$.  These effects cancel each other out.

\section{Relationship to the Weak Gravity Conjecture}

In \cite{wgc}, an upper bound on the strength of gravity relative to gauge interactions in quantum gravity was proposed.  In particular, it was conjectured that any low-energy effective Abelian gauge theory (or non-Abelian gauge theory that can be Higgsed to $U(1)$'s) that admits a consistent ultraviolet completion must contain a state with mass to charge ratio $M/q$ less than or equal to $1$.  This conjecture was motivated by the absence of black hole remnants and the non-existence of global symmetries in quantum gravity, independent of any arguments coming from string theory (though examples from string theory were used to provide further evidence for the conjecture).  The authors of \cite{wgc} noted that the observation of \cite{Banks} that axion decay constants cannot be made parametrically larger than $M_p$ without introducing additional harmonics to the potential is subsumed in the generalized `weak gravity conjecture.'  We repeat their argument here.

Consider a $p$-form Abelian gauge field in $D$ dimensions.  The natural generalization of the weak gravity conjecture for vector gauge fields holds that there should be electrically and magnetically charged $p-1$ and $D-p-1$ dimensional objects with respective tensions,
\begin{equation}
T_{\text{el}} \lesssim \left( \frac{g^2}{G_N} \right)^{1/2}\,,~~~~T_{\text{mag}} \lesssim \left( \frac{1}{g^2 G_N} \right)^{1/2}.
\label{generalized}
\end{equation}
Setting $p=0$ for a 0-form axion, we expect the theory to contain an object of tension $T_{\text{el}} \sim g$.  In this case, the object is just the instanton that couples to the axion, the tension is just the instanton action, and the gauge coupling is given by the reciprocal of the axion decay constant $g \sim 1/f$, so (\ref{generalized}) becomes,
\begin{equation}
S_{\text{inst}} \lesssim \frac{M_p}{f}.
\label{instantonaction}
\end{equation}
The axion potential is of the form,
\begin{equation}
V \sim \sum_n e^{-n S_{\text{inst}}} \cos{n\phi/f},
\end{equation}
so whenever $f$ becomes larger than $M_p$, (\ref{instantonaction}) implies that higher order terms in the sum necessarily become large.

As a further piece of evidence for the generalized weak gravity conjecture, it was pointed out that this conjecture would explain why the inflation scenario of \cite{Randall} resists an embedding into string theory.

In \cite{Cheung}, the weak gravity conjecture was extended to theories with multiple $U(1)$'s.  Interestingly, the na\"ive extension that there must exist a particle of mass to charge ratio $M/q \leq 1$ for each $U(1)$ is insufficient.  Rather, the correct extension to $N$ $U(1)$'s is that there must exist a collection of particle species $i=1,...,N$ with charge vectors $\vec{q}_i$ and charge-to-mass vectors $\vec{z}_i = \vec{q}_i \frac{M_p}{m_i}$, and the convex hull spanned by the vectors $\pm \vec{z}_i$ must contain the $N$-dimensional unit ball.  We will now show that this extension of the (generalized) weak gravity conjecture explains the difficulty of achieving parametric enhancement of axion moduli space diameters in string theory.

Consider a theory with $N$ axions of decay constants $f_1,...,f_N$.  Each of these axions couples to an instanton of action $S_{\text{inst}}^{(i)}, i = 1,...,N$.  For a 0-form axion, the appropriate analog of the charge-to-mass vector is $\vec{z}_i = \frac{g_i}{S_{\text{inst}}^{(i)}} \vec{e}_i = \frac{M_p}{f_i S_{\text{inst}}^{(i)}} \vec{e}_i$, where the $\{\vec{e}_i\}$ form an orthonormal basis of the vector space.  It is not too difficult to show that the requirement that the convex hull spanned by the vectors $\pm \vec{z}_i$ must contain the unit ball imposes the constraint, $\sum_i \frac{1}{|\vec{z}_i|^2} \leq 1$.  Thus,
\begin{equation}
\displaystyle\sum_{i=1}^N \left(f_iS_{\text{inst}}^{(i)}\right)^2 \leq 1.
\label{cond}
\end{equation}
Recalling that additional harmonics up to order $n$ are expected to become important when $n S_{\text{inst}}^{(i)} \sim 1$, we may without loss of generality take $S_{\text{inst}}^{(i)} > 1$.  The effective axion decay constant is then given by $f_{i_{\text{eff}}} \sim f_i  S_{\text{inst}}^{(i)}$, so that the condition (\ref{cond}) becomes,
\begin{equation}
\displaystyle\sum_{i=1}^N f_{i_{\text{eff}}}^2 \lesssim 1.
\end{equation}
Note that the requirement that the inflaton should have mass $\sim 10^{-5} M_p$ imposes an even stronger condition $S_{\text{inst}}^{(i)} \gg 1$ (assuming the potential is not significantly suppressed by an additional prefactor).  Hence, the effective axion decay constants scale as $1/\sqrt{N}$, negating the gain from traveling along the diagonal.  When applied to axions, the $N$-species extension of the weak gravity conjecture derived in \cite{Cheung} is precisely the statement that the diagonal of axion moduli space is bounded.  Since the weak gravity conjecture is thought to apply not only to string theory, but to any consistent theory of quantum gravity, this gives us reason to believe that large axion moduli spaces are in the swampland of whichever such theory might describe our universe.

\section{Conclusions}

We have studied in detail the axion moduli spaces of type IIB compactifications on Calabi-Yau three-folds, and have derived a rigorous bound on the axion moduli space radius within the large volume limit for Calabi-Yaus with simplicial K\"{a}hler cones.  We have seen that this bound breaks down when the Calabi-Yau has non-simplicial K\"{a}hler cone, but that additional contributions to the axion potential are introduced to the 4d effective action.  A statistical analysis of simulated Calabi-Yau moduli spaces suggests that these contributions become important whenever the radius of moduli space becomes greater than the bound (\ref{rbound2}), cutting off the range accessible to the axions and prohibiting a large traversal during inflation.  We have briefly examined axion moduli spaces of type IIA and heterotic string theory, finding further heuristic evidence that axion moduli space radii should not scale parametrically with the number of axions as $\mathcal{O}(N^{1/2})$, as has been assumed previously \cite{Nflation}, but rather as $\mathcal{O}(N^0)$.

The line of reasoning is very similar to the single axion case.  There, one cannot readily find a super-Planckian decay constant in the weakly coupled, large volume regime of parameter space.  And as soon as one tries to leave the large-volume limit, additional contributions to the potential become important and render inflation impossible.  Here, when additional axions are included, the Planck mass scales accordingly, maintaining a small moduli space radius regardless of the number of axions.  One can extend the moduli space radius by carefully inducing cancellations to lower the Calabi volume and hence decrease the Planck mass, but (at least in our simulated examples) this also comes at the expense of introducing new, important contributions to the potential that render inflation impossible.  This makes a great deal of sense from the effective field theory perspective--the absence of global symmetries in quantum gravity means that the axion's shift symmetry must be broken at high energy scales, presumably by Planck-suppressed operators.  This implies that one should not expect to find large, flat regions of the potential over distances larger than $\mathcal{O}(M_p)$ in field space.  One would not expect this principle to depend on the dimensionality of field space--large, flat distances in an $N$-dimensional field space are just as unnatural to the effective field theorist as are large, flat distances in $1$-dimensional field space.  Our analysis suggests that the effective field theorist's squabbles with super-Planckian traversals are not circumvented by high-dimensional axion moduli spaces.  Furthermore, the relationship demonstrated here between axion moduli spaces and the weak gravity conjecture strongly suggests that a parametrically large axion moduli space radius is not only in the swampland of string theory, but also is incompatible with any consistent theory of quantum gravity.

We have made a few simplifying assumptions in our statistical analysis, all of which point to possible future lines of work.  On the mathematical side, it would be very interesting to study the secondary fans of toric varieties admitting Calabi-Yau hypersurfaces and to see under what conditions and how frequently one expects the nef cone of the Calabi-Yau to be a simple union of nef cones of ambient toric varieties.  With a better understanding of the nef cones of Calabi-Yaus, one could improve on our analysis by studying true Calabi-Yau axion moduli spaces, rather than our simulated examples, and compute exactly the axion moduli space radii.  It would also be nice to understand the moduli spaces of Calabi-Yaus with extra generators of the K\"{a}hler cone not inherited from the ambient space (i.e. those for which the polytope is not favorable).  Finally, one would like a better understanding of the cone of effective divisors of Calabi-Yau three-folds, which is necessary for more detailed studies of $\vartheta_i$ axion moduli spaces.

Cosmologically, the difficulty of finding a large axion moduli space radius means that natural inflation may be extremely difficult to achieve in string theory, or any consistent theory of quantum gravity.  The arguments presented here are not rigorous enough to definitively rule out the possibility of a parametric enhancement of the axion moduli space radius.  Furthermore, multi-axion inflation does not require \emph{parametric} enhancement of the field range well beyond $M_p$ to agree with experiment--merely a field range of order $10 \, M_p$.  Nonetheless, these arguments still present difficulties from a model-building perspective, and they also give us reason to question our original motivation for the theory of multi-axion inflation: one of the primary reasons for recruiting extra axions was the discovery in \cite{Banks} of string theory's bias against super-Planckian individual axion decay constants based on heuristic arguments from different corners of the duality web.  We have now seen similar arguments against large traversals in multi-axion scenarios and have (indirectly) linked this problem to the problem of black hole remnants via the weak gravity conjecture.  It appears to us equally reasonable, therefore, to search for small $N$ counterexamples to our bound (provided the K\"{a}hler cone is non-simplicial) as it does to search for large $N$ counterexamples.  However, we suspect that both will be hard to come by, and one will have to utilize axion monodromy or else venture even deeper into the bowels of string theory to generate a large tensor-to-scalar ratio from a model of string inflation.

We wish to thank C. Vafa, D. Morrison, H. Ooguri, L. McAllister, C. Long, M. Alim, S. Katz, W. Taylor, M. Esole, M. Reece, A. Strominger, J. Harris, and T. Dumitrescu for helpful discussions.  We also wish to thank A. Chi for assistance with computer programming.  This material is based upon work supported by the National Science Foundation under Grant No. DGE-1144152.  Portions of this work were carried out at the Primordial Gravitational Waves and Cosmology Workshop (California Institute of Technology), String-Math 2014 (University of Alberta), and the 2014 Simons Summer Workshop in Mathematics and Physics (Simons Center for Geometry and Physics).  We thank all of these institutions for hospitality.

\appendix

\section{Intersection Theory and Toric Geometry}

We present a brief review of the relevant aspects of toric geometry and intersection theory.  A more detailed introduction can be found in e.g. the texts \cite{KatzCox, VafaMirrorSymmetry, CoxToricVarieties}.

An $r$-dimensional toric variety $X$ is an algebraic variety over $\mathbb{C}$ containing a complex torus $T = (\mathbb{C}^*)^r$ as a dense open set, along with an action of $T$ that extends to the entire space, $X$.  The data of a toric variety can be encoded in a fan of cones over a lattice, as a polytope in a dual lattice, or as the ground states of a gauged linear sigma model (GLSM).  We will briefly discuss the relevant aspects of each of these and present a simple example.

\begin{figure}
\begin{center}
\includegraphics[trim = 12mm 5mm 12mm 5mm, clip, width=4cm]{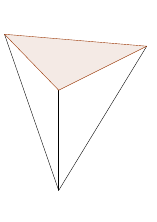}
\includegraphics[trim = 12mm 5mm 13mm 7mm, clip, width=5cm]{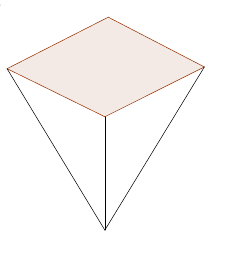}
\end{center}
\caption{A simplicial (left) cone and non-simplicial (right) cone.}
\label{nonsimplicial}
\end{figure}

Given a lattice $N \cong \mathbb{Z}^r$, set $N_{\mathbb{R}} = N \otimes \mathbb{R}$.  A strongly convex rational polyhedral cone $\sigma \subset N_{\mathbb{R}}$ is a set consisting of all positive linear combinations of a specified set of vectors in $N$.  A fan consists of a collection of strongly convex rational polyhedral cones with the property that the face of each cone is also a cone in the lattice, and the intersection of two cones in the lattice is a face of each.  If the cones span $N_\mathbb{R}$, the toric variety is compact.  The one-dimensional cones (cones generated by a single vector in $N$) of a fan are known as rays.  If the number of generating rays of a cone is equal to the dimensionality of the cone, the cone is said to be simplicial.  Otherwise, the number of generating rays must be greater than the dimensionality of the cone, and the cone is called non-simplicial, as shown in Figure \ref{nonsimplicial}.

If there are $n+r$ rays of the fan, there will be $n$ linear relations between them.  Each of these relations corresponds to a $U(1)$ charge from the GLSM perspective, with the $n+r$ rays each yielding a field $\phi_i$ with charge vector $Q_i^a$, $i =1,..,r+n$, $a=1,..,n$.  Each of the rays is further associated to a divisor of $X$.  As we will see, the intersections of these divisors may be easily computed from the GLSM approach.\\ \noindent
\\ \noindent
\textit{Example: Hirzebruch surface $\mathbb{F}_n$.}
\\ \noindent
A Hirzebruch surface is the $\mathbb{P}^1$ bundle over $\mathbb{P}^1$ associated to the projectivized line bundle $O(0) + O(-n)$.  It is represented by the 2d fan spanned by the vectors $v_1 = (1,0)$, $v_2 = (0,1)$, $v_3 = (0,-1)$, $v_4 = (-1,-n)$.  In 2d, a fan is specified uniquely by its rays, so this completely defines the toric variety.  There are two relations between the vectors of the fan:
\begin{equation}
v_2 + v_3 = 0\,,~~~~ v_1 + n v_2 + v_4 =0.
\end{equation}
This leads to charge vectors $(0,1,1,0)$ $(1,n,0,1)$ and hence D-term equations for the GLSM,
\begin{equation}
|\phi_2|^2 + |\phi_3|^2 = r_1\,,~~~~|\phi_1|^2 + n |\phi_2|^2 + |\phi_4|^2 = r_2,
\label{Dterm}
\end{equation}
with a $U(1)$ action acting on each of the fields according to its corresponding charges, $\phi_i \rightarrow e^{2 \pi i Q_i^a} \phi_i$.

\hfill $\square$

Given an algebraic variety, the set of positive linear combinations of (real homology classes of) irreducible, reduced, proper curves forms a cone known as the Mori cone.  For a toric variety $X$ with fan $\Sigma$, the generators of this cone, which are the classes of holomorphic curves in $X$, correspond to $(r-1)$-dimensional cones in $\Sigma$.

Each ray in $\Sigma$ corresponds to a divisor in $X$.  A divisor $D$ is called nef if it satisfies $D \cdot C \geq 0$ for every irreducible curve $C$ (in particular, for any element in the Mori cone).  The collection of such nef divisors also forms a cone, which is known as the nef cone.  The K\"{a}hler cone of $X$ is the interior of the nef cone.  Faces of the nef cone correspond to generating curves of the Mori cone, and faces of the Mori cone correspond to generating divisors of the nef cone in the sense that a curve will shrink at the face of the nef cone, while a divisor will shrink at a face of the Mori cone.

The nef cone and its interior, the K\"{a}hler cone, will be of particular interest for our purposes.  It may be viewed as the intersection of all cones complementary to the $r$-dimensional cones of $\Sigma$.  In other words, given an $r$-dimensional cone in $\Sigma$, spanned by rays $v_i, i \in I$, we consider the cone spanned by the corresponding charge vectors $Q_i^a, i \in \{1,..,n+r\} \setminus I$, the complement of $I$.  We label this cone a `charge cone,' and the intersection over all such charge cones associated to the toric variety is the nef cone of $X$, which we will sometimes denote Nef($X$).  The collection of all GKZ cones forms a fan known as the secondary fan.

A basis of divisors in $X$ is given by the the set $\{ D_i \}$, where $D_i$ is the vanishing locus of $\phi_i$, a field of the GLSM.  The intersection number of $r$ divisors $\{D_{i_1},...D_{i_r}\}$ in $X$ can further be computed from the GLSM description, and is given (roughly) by the number of solutions to the D-term equations when each of the $\{ \phi_{i_k} \}$ is set to zero.  Given a gauge invariant combination $\phi_{i_1}^{n_1} \cdot...\cdot \phi_{i_m}^{n_m}, n_k \in \mathbb{Z}$, there is an equivalence relation $n_1 D_{i_1} + ... + n_m D_{i_m} = 0$ that commutes with the intersection product operation.  This can be used to define formal self-intersection numbers such as $D_i^r$.

All intersection numbers of a collection of nef divisors are non-negative \cite{positivity}.
\\ \noindent
\textit{Example: Hirzebruch surface $\mathbb{F}_n$.}
\\ \noindent
We return to our example of a Hirzebruch surface.  There are four 2-dimensional cones in the fan of $\mathbb{F}_n$, each composed of two adjacent 1-cones.  Writing the GLSM charges in matrix form for purposes of illustration,
\begin{equation}
\left( \vec{\beta}_1 \, \vec{\beta}_2 \,  \vec{\beta}_3 \,  \vec{\beta}_4 \right) = \left( \begin{array}{cccc}
1 & n & 0 & 1 \\
0 & 1 & 1 & 0 \end{array} \right),
\end{equation}
and the fan in matrix form
\begin{equation}
\left( \vec{v}_1 \, \vec{v}_2 \,  \vec{v}_3 \,  \vec{v}_4 \right) = \left( \begin{array}{cccc}
1 & 0 & 0 & -1 \\
0 & 1 & -1 & -n \end{array} \right),
\end{equation}
the 2-dimensional cones of the fan are generated respectively by $\{\vec{v}_1,\vec{v}_2\}$, $\{\vec{v}_2,\vec{v}_4\}$, $\{\vec{v}_4,\vec{v}_3\}$, $\{\vec{v}_3,\vec{v}_1\}$.  The GKZ cone for this fan, and hence its nef cone, is given by the intersection of the charge cones generated by the complementary sets of vectors, $\{\vec{\beta}_3,\vec{\beta}_4\}$, $\{\vec{\beta}_1,\vec{\beta}_3\}$, $\{\vec{\beta}_1,\vec{\beta}_2\}$, $\{\vec{\beta}_2,\vec{\beta}_4\}$.  This is the region in the first quadrant spanned by $\{\vec{\beta}_1,\vec{\beta}_2\}$.  The intersection numbers of the divisors $D_1$, $D_2$ spanning the nef cone can be determined from the GLSM.  The divisors satisfy equivalence relations $D_1 = D_4$, $D_2 = n D_1 + D_3$.  Setting $\phi_2 = \phi_4 = 0$, we find precisely one solution to the D-term equations (\ref{Dterm}), so $D_2 \cdot D_4 = 1$.  Likewise, setting $\phi_2 = \phi_3 = 0$, we find no solutions to the second D-term equation, so $D_2 \cdot D_3 = 0$.  Setting $\phi_3 =\phi_4 = 0$ gives $D_3 \cdot D_4 = 1$.  This suffices to fix the intersection numbers of all the divisors.  In particular, for the generators $D_1$ and $D_2$ of the nef cone,
\begin{equation}
D_1^2 = 0,\,~~~~D_1 \cdot D_2 =1 ,\,~~~~ D_2^2 = n.
\end{equation}
Note that these intersection numbers are all non-negative, as they must be for nef divisors.

 \hfill $\square$

The data of a toric variety can also be encoded in a lattice polytope $\Delta$, which is the convex hull in $\mathbb{R}^r$ of a set of lattice points of the lattice $M$, which is dual to the lattice $N$.  Restricting to normal varieties, we may define a fan directly from the polytope as follows: for each face $F$ of the polytope, define a cone $\sigma_F$ by
\begin{equation}
\sigma_F = \{ \vec{v} \in N_{\mathbb{R}} | \langle \vec{m}, \vec{v} \rangle \leq \langle \langle \vec{m}', \vec{v} \rangle \mbox{ for all } \vec{m} \in F, \vec{m}' \in \Delta\}.
\end{equation}
The collection of all such cones $\sigma_F$ forms a fan known as the normal fan.  The toric variety associated with the normal fan is thus associated with the polytope $\Delta$.
\\ \noindent
\textit{Example: Hirzebruch surface $\mathbb{F}_n$.}
\\ \noindent
Returning once again to the Hirzebruch surface, we find that the polytope,
\begin{equation}
\Delta = \mbox{Hull}\{(0,0),(0,1),(1,1),(n,0)\}
\end{equation}
generates the toric variety $\mathbb{F}_n$.  It is easy to see that the normal fan of this polytope is the fan of $\mathbb{F}_n$ considered previously.

\hfill $\square$

A polytope $\Delta$ is called reflexive if its only interior lattice point is the origin.  In such a case, the generating rays of the normal fan will be integral (i.e. will lie in $N$ rather than just $N_{\mathbb{R}}$), and we may define a dual lattice polytope $\Delta^\circ$ whose vertices are the generating rays of the normal fan.  Further, $\Delta^\circ$ will also be reflexive.

In dimension greater than 2, there can be multiple fans with the same set of rays, depending on how these rays are grouped into cones.  This corresponds to a triangulation of the dual polytope $\Delta^\circ$ i.e. a different grouping of the vertices into $r-1$-simplices (i.e. facets of the polytope).  Each simplex gives rise to a different charge cone, so each triangulation gives rise to a different nef cone (which is the intersection of all such charge cones for the $r-1$ simplices of the polytope).  Adjacent nef cones correspond to toric varieties that are related by birational transformations, which are maps that are isomorphisms except on subsets of codimension one or greater.  This is depicted beautifully in Figure 6 of Chapter 15 of \cite{CoxToricVarieties}.

Finally, we consider Calabi-Yaus embedded in toric varieties.  A compact toric variety itself will not be Calabi-Yau.  Instead, we must consider a subvariety of the toric variety with vanishing first Chern class.  More precisely, we take a reflexive polytope and look at its associated toric variety $X$.  We then consider an anticanonical hypersurface $Z \subset X$, which is a generic section in $\Gamma(X, \mathcal{O}_X(-K_X))$.  Intersections of divisors of $X$ with the Calabi-Yau hypersurface $Z$ are themselves divisors of $Z$.  Assuming that all of the divisors of $Z$ are inherited from $X$ in this way, the intersection numbers of the Calabi-Yau are similarly inherited by taking $D_i^Z \cdot ... \cdot D_j^Z = (-K_X) \cdot D_i^X \cdot ... \cdot D_j^X$.  In this situation, the polytope of $X$ is said to be favorable.  

To get Calabi-Yau manifolds with the properties we want to study, we must place restrictions on the triangulations of $\Delta^\circ$.  First off, to guarantee the existence of a Calabi-Yau manifold, we insist that the toric variety should be Gorenstein and Fano, which means that $\Delta^\circ$ must be a reflexive lattice polytope.  Next, we insist that the triangulation should be star, which simply means that all simplices should contain the origin, so that all cones in the fan are generated by rays emanating from the origin.  To ensure that the K\"{a}hler cone of the toric variety is nonempty, we further insist that the triangulations of $\Delta^\circ$ should be regular.  This means that if the points of the triangulation are labeled as $\{\vec{v}_k\}$, then there must exist an $(r+1)$-dimensional polytope with points $\{(\vec{v}_k,u_k)\}$ whose outward-facing normals have negative $(r+1)$-dimensional component.  For the purposes of computing the intersections of divisors, we restrict to favorable polytopes.  The favorableness of a polytope can be readily determined using the construction in \cite{Batyrev}.

To get a smooth toric variety, one must consider only complete triangulations--that is, triangulations in which all points in $\Delta^\circ \cap N$ are vertices of some simplex of the triangulation.  In the aforementioned Figure 6 in Chapter 15 of \cite{CoxToricVarieties}, for instance, the three triangulations shown in the first quadrant are complete, whereas the other two triangulations are not.  Smoothness of the toric variety is not necessary for smoothness of the Calabi-Yau hypersurface, which will generically avoid singular points of the ambient toric variety.  Such singularities are encoded by triangulations in which all points in $\Delta^\circ \cap N$ \emph{except} those in the interior of codimension one faces of the polytope.  Thus, to get smoothness, we need not worry about points interior to facets in our triangulations.

In fact, smoothness of the Calabi-Yau is itself not necessary to produce an acceptable 4d field theory upon compactification.  As discussed in \cite{WittenMF}, there will also be singular phases in type II compactifications.  However, the thickness of the regions of such phases is suppressed by $\alpha'$, and so these regions disappear in the limit $\alpha' \rightarrow 0$ \cite{MorrisoncitedinWitten}, which is precisely the large volume limit in which we are interested.  Thus, we will restrict our attention to smooth Calabi-Yau manifolds in the present study.

\begin{figure}
\begin{center}
\includegraphics[trim = 5mm 5mm 13mm 8mm, clip, width=7cm]{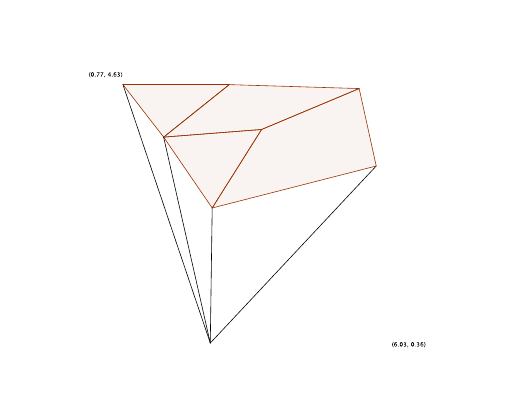}
\end{center}
\caption{The nef cone of a Calabi-Yau hypersurface typically contains the union of the nef cones of multiple ambient toric varieties related by birational transformations.}
\label{multipleconepicture}
\end{figure}

The nef cone of $Z$, Nef($Z$) will typically contain the union of the nef cones of many ambient toric varieties, related to each other by phase transitions.  This can be understood as follows: consider a Calabi-Yau hypersurface $Z$ in one particular ambient toric variety, $X$.  $X$ admits a nef cone of divisors which pullback to nef divisors on $Z$.  As we approach a wall of Nef($X$), a curve in $X$ will shrink to zero size.  If this curve intersects the hypersurface $Z$, then $Z$ will also see a curve shrinking to zero size, and correspondingly this wall serves as a boundary of Nef($Z$) as well as Nef($X$).  But if, on the other hand, the curve does not intersect $Z$ (i.e. if its intersection number with the anticanonical divisor $- K_X$ vanishes), then the shrinking of the curve will be invisible to $Z$--the nef cone of $Z$ will continue past the wall into a new GKZ chamber--the nef cone of a different ambient toric variety $\tilde{X}$ that is related to $X$ by a birational transformation called a 'trivial flop.'  We refer to such walls as `paper walls,' to distinguish them from the `solid walls' which are boundaries of both Nef($X$) and Nef($Z$).  Nef($Z$) will then contain the union of the nef cones of both $X$ and $\tilde{X}$, and indeed will contain the nef cones of all toric varieties related to $X$ by a sequence of trivial flops, as depicted in Figure \ref{multipleconepicture}.

\begin{figure}
\begin{center}
\includegraphics[width=100mm]{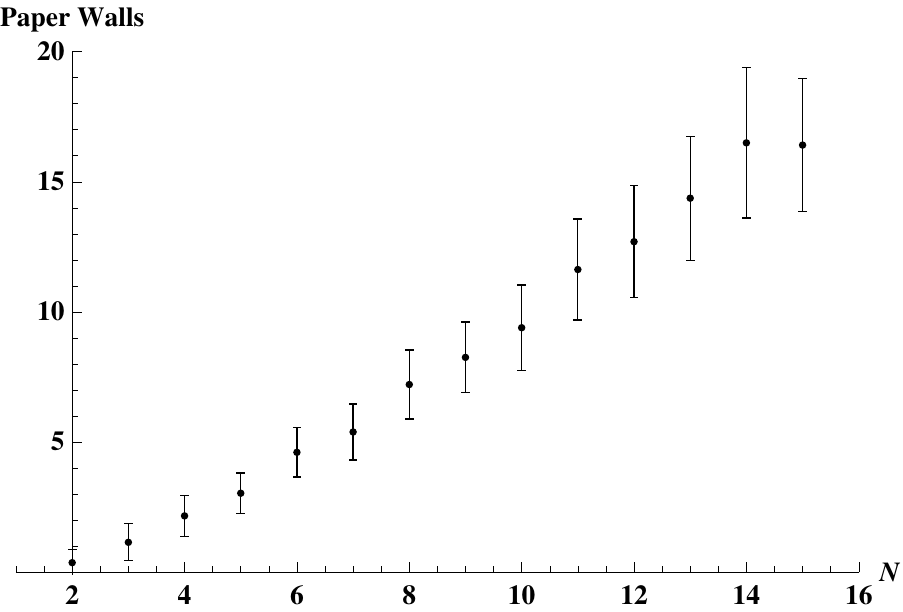}
\end{center}
\caption{The number of paper walls, indicating the number of toric varieties related to $X$ by a single trivial flop, whose GKZ chambers are adjacent to that of $X$.  The data suggest exponential growth in the number of toric varieties related to $X$ by sequences of trivial flops as a function of $N$.}
\label{paperwallsplot}
\end{figure}

In fact, as demonstrated in \cite{CoxKatzCounterexample}, Nef($Z$) can be even larger than the union of all such cones, so that even a 'solid wall' of Nef($X$) is not necessarily a wall of Nef($Z$).  This additional enhancement is nontrivial to detect, and makes the K\"{a}hler cone of the Calabi-Yau impossible to compute by current methods.  We do not know how frequently such enhancement occurs and so cannot justify ignoring this complication.  Furthermore, since the number of trivial flops of $X$ is observed to grow linearly with $h^{1,1}=N$ (c.f. Figure \ref{paperwallsplot}), the number of sequences of trivial flops and hence the number of GKZ chambers restricting to Nef($Z$) grows exponentially with $N$.  Coupled with the increasing computational power needed to compute each individual GKZ chamber with increasing $N$, this makes statistical studies of even those Calabi-Yaus not exhibiting the enhancement phenomenon (whichever they might be) extremely difficult.  Nef($Z$) can also be far more complicated than the situation we are considering here \cite{Wilson, MorrisonMirrorSymmetry}.  

The full K\"{a}hler moduli space is composed of the union of many such nef cones related by birational transformations across solid walls \cite{BeyondKahlerCone}.  For this study of axions, however, we simply fix the K\"{a}hler moduli within one particular nef cone, restricting ourselves to one geometric phase of the Calabi-Yau.

\begin{figure}
\begin{center}
\includegraphics[trim = 5mm 12mm 13mm 12mm, clip, width=8cm]{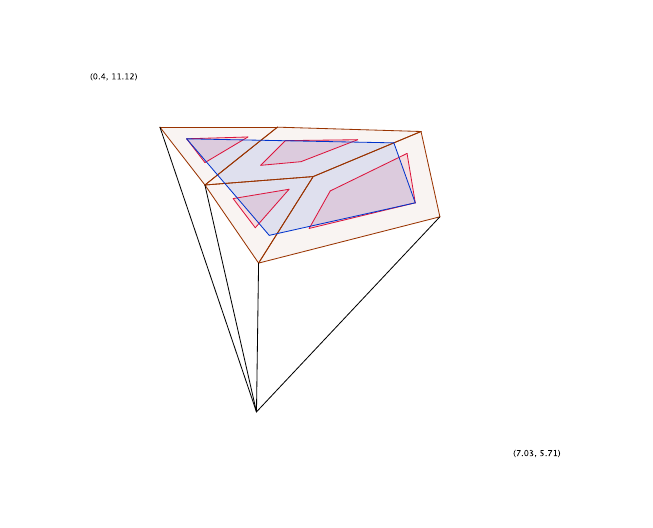}
\end{center}
\caption{The large volume regions of parameter space of the toric varieties (shown in red) are distinct from those of the entire Calabi-Yau (shown in blue).}
\label{LVSpicture}
\end{figure}

To stay in the large volume limit, we require the volumes of all irreducible effective curves to be $\gtrsim \alpha'$.  In particular, this means that non-perturbative corrections to the metric will be exponentially suppressed, and so can be ignored.  Geometrically, this means that we must stabilize K\"{a}hler moduli away from the walls of Nef($Z$), where some curve is shrinking to zero size.  As shown in Figure \ref{LVSpicture}, setting all curves of the ambient space to have volume $\gtrsim \alpha'$ is neither necessary nor sufficient to stay in the large volume limit of the Calabi-Yau itself.  Clearly, one can approach one of the paper walls, leaving the large volume limit of the ambient toric variety (shown in red in Figure \ref{LVSpicture}), but remaining in the large volume limit of the Calabi-Yau (shown in blue).  Furthermore, although not as obvious, the large volume limit of the ambient space is not necessarily contained in the large volume limit of the Calabi-Yau.  This might be thought of intuitively as a difference in metrics on the two nef cones: being some distance $d$ away from a wall of Nef($X$) is not the same as being $d$ away from a wall of Nef($Z$).

\end{document}